\documentclass[aps,prx,longbibliography,10pt,amsmath,amssymb,twocolumn,footinbib,superscriptaddress]{revtex4-1}
\usepackage{color, braket}
\usepackage[large,bf]{subfigure}     
\usepackage[breaklinks]{hyperref}
\usepackage{dsfont}

\definecolor{purple}{rgb}{0.5,0,0.5}
\definecolor{dkgreen}{rgb}{0,0.5,0}
\definecolor{orange}{rgb}{1,0.5,0}

\usepackage{bbm}

\usepackage{ifpdf}
\ifpdf
    \usepackage{psfrag}
    \usepackage[pdftex]{epsfig}
\else
    \usepackage{psfrag}
    \usepackage[dvips]{epsfig}
\fi

\begin{document}

\title{Milestones toward Majorana-based quantum computing}
\date{\today}
\author{David Aasen}
\thanks{These authors contributed equally to this work.}
\affiliation{Department of Physics and Institute for Quantum Information and Matter, California Institute of Technology, Pasadena, CA 91125, USA}
\author{Michael Hell}
\thanks{These authors contributed equally to this work.}
\affiliation{Division of Solid State Physics and NanoLund, Lund University, Lund, Sweden}
\affiliation{Center for Quantum Devices and Station Q Copenhagen, Niels Bohr Institute, University of Copenhagen, Copenhagen, Denmark}
\author{Ryan V. Mishmash}
\thanks{These authors contributed equally to this work.}
\affiliation{Department of Physics and Institute for Quantum Information and Matter, California Institute of Technology, Pasadena, CA 91125, USA}
\affiliation{Walter Burke Institute for Theoretical Physics, California Institute of Technology, Pasadena, CA 91125, USA}
\author{Andrew Higginbotham}
\thanks{These authors contributed equally to this work.}
\affiliation{Department of Physics, Harvard University, Cambridge, Massachusetts, 02138, USA}
\affiliation{Center for Quantum Devices and Station Q Copenhagen, Niels Bohr Institute, University of Copenhagen, Copenhagen, Denmark}
\author{Jeroen Danon}
\affiliation{Center for Quantum Devices and Station Q Copenhagen, Niels Bohr Institute, University of Copenhagen, Copenhagen, Denmark}
\affiliation{Niels Bohr International Academy, Niels Bohr Institute, University of Copenhagen, Copenhagen, Denmark}
\author{Martin Leijnse}
\affiliation{Division of Solid State Physics and NanoLund, Lund University, Lund, Sweden}
\affiliation{Center for Quantum Devices and Station Q Copenhagen, Niels Bohr Institute, University of Copenhagen, Copenhagen, Denmark}
\author{Thomas S. Jespersen}
\affiliation{Center for Quantum Devices and Station Q Copenhagen, Niels Bohr Institute, University of Copenhagen, Copenhagen, Denmark}
\author{Joshua A. Folk}
\affiliation{Center for Quantum Devices and Station Q Copenhagen, Niels Bohr Institute, University of Copenhagen, Copenhagen, Denmark}
\affiliation{Quantum Matter Institute, University of British Columbia, Vancouver, British Columbia, V6T1Z4, Canada}
\affiliation{Department of Physics and Astronomy, University of British Columbia, Vancouver, British Columbia, V6T1Z4, Canada}
\author{Charles M. Marcus}
\affiliation{Center for Quantum Devices and Station Q Copenhagen, Niels Bohr Institute, University of Copenhagen, Copenhagen, Denmark}
\author{Karsten Flensberg}
\affiliation{Center for Quantum Devices and Station Q Copenhagen, Niels Bohr Institute, University of Copenhagen, Copenhagen, Denmark}
\author{Jason Alicea}
\affiliation{Department of Physics and Institute for Quantum Information and Matter, California Institute of Technology, Pasadena, CA 91125, USA}
\affiliation{Walter Burke Institute for Theoretical Physics, California Institute of Technology, Pasadena, CA 91125, USA}


\begin{abstract}
We introduce a scheme for preparation, manipulation, and readout of Majorana zero modes in semiconducting wires with mesoscopic superconducting islands.  Our approach synthesizes recent advances in materials growth with tools commonly used in quantum-dot experiments, including gate-control of tunnel barriers and Coulomb effects, charge sensing, and charge pumping.  We outline a sequence of milestones interpolating between zero-mode detection and quantum computing that includes (1) detection of fusion rules for non-Abelian anyons using either proximal charge sensors or pumped current; (2) validation of a prototype topological qubit; and (3) demonstration of non-Abelian statistics by braiding in a branched geometry.  The first two milestones require only a single wire with two islands, and additionally enable sensitive measurements of the system's excitation gap, quasiparticle poisoning rates, residual Majorana zero-mode splittings, and topological-qubit coherence times.  These pre-braiding experiments can be adapted to other manipulation and readout schemes as well.

\end{abstract}

\maketitle

\section{Introduction}
\label{Introduction}

Over the span of a few years, Majorana zero modes in topological superconductors evolved from a largely theoretical topic into an active experimental field at the forefront of condensed matter physics \cite{BeenakkerReview,AliceaReview,FlensbergReview,TewariReview,FranzReview,NayakReview}.  This transformation was driven in part by the translation of abstract models \cite{ReadGreen,1DwiresKitaev} into realistic blueprints realizable with established laboratory capabilities (see, e.g., Refs.~\onlinecite{FuKane,MajoranaQSHedge,Sau,Alicea,1DwiresLutchyn,1DwiresOreg,NoSOC1,Yazdani}). Proposals based on semiconductor nanowires \cite{1DwiresLutchyn,1DwiresOreg} provide a good illustration of this evolution.  Within this approach, stabilizing Majorana zero modes requires spin-orbit coupling, moderate magnetic fields, and proximity-induced superconductivity.

Following the initial experiments of Mourik et al.~\cite{mourik12} in 2012, several groups reported transport features consistent with Majorana modes in a variety of related superconductor-semiconductor systems \cite{das12,Rokhinson,deng12,finck12,Churchill} (see also Ref.~\onlinecite{Nadj-Perge}).  In parallel, fabrication advances \cite{Gul,Marcus15_NatureMat_14_400} have improved device quality, leading to cleaner transport characteristics \cite{Marcus15_NatureNano_10_232} as well as surprisingly long quasiparticle poisoning times in proximitized nanowires \cite{PoisoningTime} and related setups \cite{Woerk}.  Branched wires, another key ingredient toward Majorana networks \cite{AliceaBraiding,ClarkeBraiding,SauBraiding,HalperinBraiding,BraidingWithoutTransport,BeenakkerBraiding,BondersonBraiding}, have also been realized and investigated recently \cite{WireNetworks}.

With this rapid progress, we anticipate that nanowire-based experiments will soon move beyond the problem of Majorana detection to demonstrations of non-Abelian statistics and ultimately to implementations of topological quantum-information processing \cite{kitaev,TQCreview,NayakReview}.  Our objective in this paper is to facilitate this progression in two ways.  First, we propose a new method for manipulation and readout of Majorana modes in semiconducting wires that borrows from the quantum dot toolbox. Second, within this scheme we outline a series of milestones that identify the key challenges and signature results that connect the current status of experiments to these longer-term challenges.

\begin{figure}
\includegraphics[width=\columnwidth]{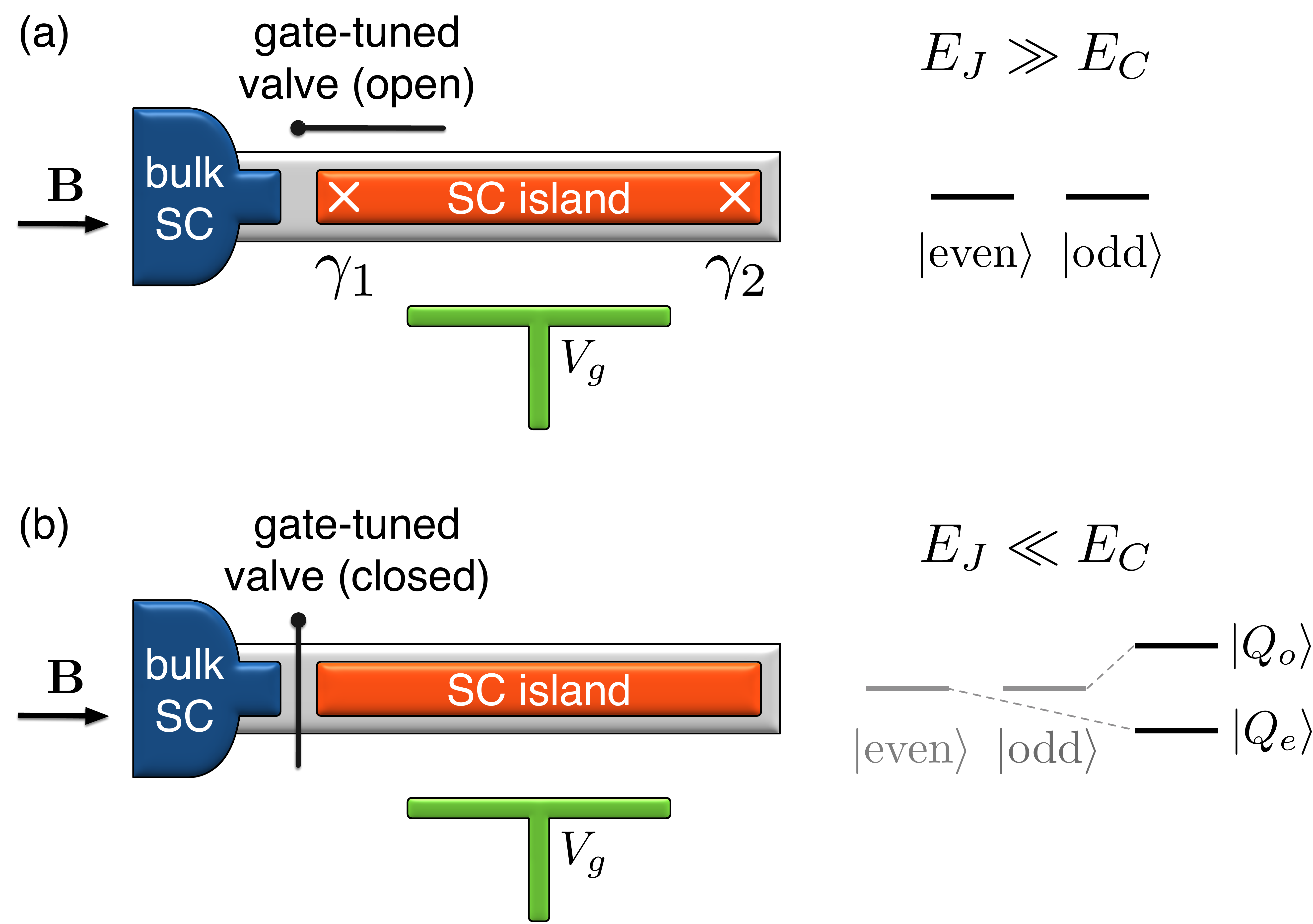}
\caption{Semiconducting wire coated with a superconducting island and bulk superconductor that are bridged by a gate-tunable `valve'.  The valve controls the carrier density in the barrier region and thereby modulates the ratio of Josephson energy $E_J$ to the island charging energy $E_C$.  In addition, a back gate at voltage $V_g$ tunes the charge on the island.  (a) When the valve is open ($E_J \gg E_C$) an applied magnetic field ${\bf B}$ drives the wire into a topological superconducting state hosting Majorana zero modes $\gamma_{1,2}$.  The system thus supports degenerate ground states with even and odd fermion parity.   (b) Closing the valve ($E_J \ll E_C$) restores charging energy and converts these parity eigenstates into non-degenerate states with island charges $Q_o$ and $Q_e$ as shown on the right.  The gate-controlled `parity-to-charge conversion' illustrated here is central to the manipulation and readout schemes developed in this paper.}
\label{SetupFig}
\end{figure}

The rudiments of our manipulation scheme can be understood from the schematic setup shown in Fig.~\ref{SetupFig}.  Here a semiconducting wire is partially coated with a mesoscopic superconducting island on the right side and a bulk grounded superconductor on the left.  The superconductors are separated by a gate-tunable `valve' that controls the ratio of the Josephson energy $E_J$ to the island's charging energy $E_C$. Additionally, the wire is subjected to a magnetic field needed to form Majorana zero modes. References \onlinecite{Gatemon,LangeNanowireJosephson} introduced a very similar device to construct a gate-controlled transmon qubit, though here we are interested in a different regime \footnote{Recently Ref.{~\onlinecite{Shim}} also utilized gate-tuned Josephson junctions to propose (non-topological) quantum computing schemes based on semiconductor-superconductor hybrids.}.
When the valve is `open'---i.e., one or more modes in the barrier provide sufficient coupling that the junction's Josephson energy dominates over charging energy---the island hosts a pair of Majorana zero modes $\gamma_{1,2}$ that encode a topological degeneracy between ground states with even and odd fermion parity; see Fig.~\ref{SetupFig}(a).  Closing the valve by depleting the barrier renders charging energy dominant over the Josephson energy, lifting the degeneracy of parity eigenstates by converting them into charge states of the island, as sketched in Fig.~\ref{SetupFig}(b).  We note that this scheme closely resembles the `Coulomb-assisted' Majorana manipulation approaches described in Refs.~\onlinecite{TopTransmon,BraidingWithoutTransport,BeenakkerBraiding} (see also Ref.~\onlinecite{MajoranaTransportWithInteractions1}), but uses depletion gates rather than fluxes to control the ratio of Josephson coupling to charging energy.

Gate tuning of Josephson and Coulomb energies is particularly useful when integrated into devices with multiple islands.  For instance, this capability allows Majorana zero modes to be initialized by evolving known charge states into degenerate parity eigenstates [e.g., smoothly passing from Fig.~\ref{SetupFig}(b) to (a)], alter the length of a topological superconductor by selectively opening and closing valves connecting adjacent islands \cite{PhaseTransition}, introduce rotations within the ground-state subspace, and braid Majorana modes in network geometries.  Converting parity eigenstates back into charge states---an adaptation of `spin-to-charge conversion' used in spin qubits---further enables readout of the state formed by Majorana modes.

We discuss three specific experiments that use these capabilities:

{\bf Fusion-rule detection (Sec.~\ref{FusionRuleSec}).}  A fundamental property of non-Abelian anyons is their behavior under fusion, which describes how these emergent particles coalesce.  In our setups, the topological superconductors' endpoints, where Majorana zero modes localize, essentially realize `Ising' non-Abelian anyons.  Ising anyons obey a particularly simple fusion rule: pairs can either annihilate or combine into a fermion $\psi$.  These two `fusion channels' correspond to the ordinary fermionic state arising from a pair of hybridized Majorana modes being empty or filled.
The presence of multiple fusion channels intimately relates to non-Abelian statistics, and in fact is commonly used to define non-Abelian anyons in the first place.  Detecting this foundational property has nevertheless received very little attention.

We introduce two experimental methods for probing Ising-anyon fusion rules, both invoking a relatively simple single-wire geometry with two superconducting islands and three gate-tunable valves.  The first method operates the valves to nucleate a set of anyons and then restore charging energy to fuse them in a manner that accesses the fusion channels with known probabilities; charge sensors detect the fusion outcomes.  The second method converts the microscopic difference between the two fusion channels into a macroscopic current, simplifying the setup by eliminating charge sensing.  This approach uses a Majorana-mediated charge pump to cyclically create and fuse anyons, shuttling a Cooper pair across the system whenever the $\psi$ channel appears.
As a bonus, the protocols permit direct measurement of additional topological data for the underlying anyon theory (see Appendix \ref{AnyonAppendix}) \footnote{We thank Nick Read for emphasizing this point to us.}, as well as time-domain measurements of device parameters such as quasiparticle poisoning times, excitation gaps, and residual Majorana-zero-mode splittings.

{\bf Topological qubit validation (Sec.~\ref{QubitSec}).}  A single-wire, two-island geometry supporting four Majorana modes realizes a prototype topological qubit.  The two fixed-parity degenerate ground states available in such a setup form the logical $|0\rangle$ and $|1\rangle$ states.  How can one validate the topological nature of this qubit (assuming the usual conditions required for topological protection are maintained)? For instance, what set of measurements will distinguish a Majorana qubit from a similar setup in which degenerate ground states instead arise from accidental zero-energy Andreev bound states? We will show that the difference can be found in the coherence times and oscillation frequency, $\omega_0$, of the qubit.  In particular, relaxation and dephasing times,  $T_1$ and $T_2$, as well as $\omega_0$,  exhibit exponential dependence on the (experimentally tunable) splitting of ground states encoded by Majorana modes, leading to scaling relations among these quantities that can be used to identify topological protection.   Topological qubit validation along these lines uses the same setup as fusion-rule detection, together with the ability to implement qubit rotations via pulsing of gate-tunable valves.

{\bf Non-Abelian statistics (Sec.~\ref{BraidingSec}).}  Moving to branched geometries, we present a new approach to demonstrating non-Abelian statistics and the associated fault-tolerant qubit rotations in wire networks. Operationally, this approach resembles Coulomb-assisted braiding \cite{BraidingWithoutTransport,BeenakkerBraiding}, though using gate voltages instead of magnetic fluxes.  The hallmark of non-Abelian statistics is that such exchanges rotate the system's quantum state within the degenerate ground-state manifold generated by the zero modes; this can be detected using either charge sensors or the Majorana-mediated pump introduced earlier. Conveniently, aside from the need for trijunctions, braiding in the setups that we study involves the same capabilities required for fusion-rule detection.

This list of goals aims to demonstrate consequences of Majorana zero modes that are directly relevant for long-term quantum computing applications.  Fusion-rule detection and topological qubit validation stand out as attractive near-term milestones that reveal important aspects of non-Abelian anyons along with useful device characteristics.  Both classes of experiments should be adaptable to other Majorana platforms and manipulation schemes as natural precursors to braiding in comparatively simple geometries.  Sec.~\ref{OutlookSec} summarizes near-term prospects and elaborates on several longer-term questions raised by our work.

\section{Building blocks}
\label{GateSec}

In this section we describe in greater depth the building blocks underlying the manipulation and readout schemes used throughout the paper.  The backbone of all devices is a clean \footnote{In the clean limit, Majorana modes exist without additional spurious low-lying states.  At least in related setups, however, localized sub-gap states need not severely limit the requirements for quantum information processing; see Ref.~\onlinecite{AkhmerovMiniGap}.} spin-orbit-coupled semiconducting nanowire subjected to both an applied magnetic field and proximity-induced superconductivity---which together allow the formation of a topological phase supporting Majorana zero modes \cite{1DwiresLutchyn,1DwiresOreg}.  More precisely, we use \emph{mesoscopic} superconductors to proximitize selected regions (islands).  The islands contain a macroscopic number of electrons but are sufficiently small that charging energies exceed temperature in the Coulomb-dominated regime.  At the valve positions, the proximitizing metal is interrupted, allowing depletion via nearby electrostatic gates.

\subsection{Single-island geometry and parity-to-charge conversion}
\label{SingleIslandSec}

\begin{figure*}
\includegraphics[width=2\columnwidth]{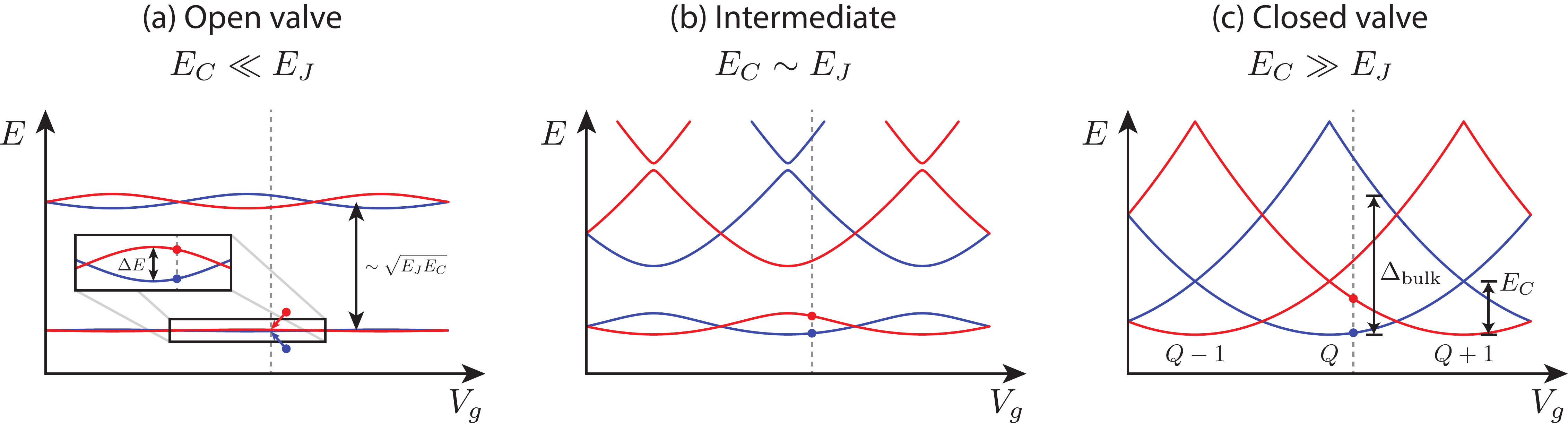}
\caption{Energy levels versus back-gate voltage $V_g$ for the single-island setup in Fig.~\ref{SetupFig} with the gate-tunable valve (a) fully open, (b) partially transmitting, and (c) fully closed.  Red and blue curves correspond to states with even and odd electron number.  For the open configuration in panel (a) Josephson energy $E_J$ dominates charging energy $E_C$.  The island thus hosts a pair of Majorana zero modes that encode an approximate two-fold topological ground-state degeneracy.  Residual charging energy splits the degeneracy, but only by an amount $\Delta E$ exponentially small in $E_J/E_C$.  Closing the valve effectively ramps up Coulomb effects, which dominate in the fully closed configuration of panel (c).  There the eigenstates are generally non-degenerate (except with fine-tuning) and carry well-defined island charges $\ldots Q-1, Q, Q+1 \ldots$ as labeled on the right.  Crucially, the topologically degenerate ground states smoothly evolve into non-degenerate charge eigenstates as the valve closes---signifying parity-to-charge conversion.  The red and blue dots illustrate this phenomenon at one particular gate voltage indicated by the vertical dashed line.  }
\label{EnergyLevels}
\end{figure*}

Let us revisit the setup in Fig.~\ref{SetupFig}. The mesoscopic island on the right forms a Josephson junction, bridged by the intervening nanowire segment, with the grounded bulk superconductor on the left.  We assume that beneath the island the nanowire chemical potential is adjusted via gates to a value required for topological superconductivity, while elsewhere a trivial gapped phase always forms.  Under this assumption, we model the  low-energy properties of the junction using a phenomenological Hamiltonian $H = H_C + H_J$.  The first term,
\begin{equation}
  H_C = E_C(\hat{n}-n_0)^2,
  \label{HC}
\end{equation}
describes Coulomb effects for the island, with $E_C$ the characteristic charging energy, $\hat{n}$ an integer-valued operator giving the electron occupancy of the island, and $n_0$ a continuous offset charge, tunable via a nearby electrostatic gate at voltage $V_g\propto n_0$ (see Fig.~\ref{SetupFig}).

The second term encodes Josephson coupling.  Assuming that the barrier hosts $N$ channels with transmission probabilities $T_{i = 1,...,N}$ and that the junction is short on the scale of the superconducting coherence length, we model the Josephson energy with a Hamiltonian
\begin{equation}
  H_J = -\Delta \sum_{i = 1}^N \sqrt{1 - T_i \sin^2 \left(\hat\varphi / 2 \right)}.
  \label{eq:hj-gen}
\end{equation}
Here $\Delta$ denotes the pairing gap (assumed equal on both sides of the valve) that sets a characteristic Josephson energy $E_J \equiv \frac{\Delta}{4} \sum_{i = 1}^N T_i$, while the operator $e^{i\hat\varphi}$ tunnels a Cooper pair across the junction, incrementing $\hat{n}$ by \emph{two}.  Equation~\eqref{eq:hj-gen} recovers the well-known form of the Josephson energy that one can derive in the absence of charging energy \cite{NazarovBook}; see also Ref.~\onlinecite{Ioselevich} for a recent discussion in a similar context.  But is $H_J$ valid also with non-zero $E_C$?  In the limit $E_J\gg E_C$ one can safely include the charging energy in Eq.~\eqref{HC} as a perturbation to Eq.~\eqref{eq:hj-gen}.  When all $T_i \ll 1$---corresponding to the opposite limit $E_C\gg E_J$ when $N$ is not too large (see below)---the Josephson Hamiltonian reduces to
\begin{equation}
  H_J \approx -E_J\cos \hat\varphi,
  \label{HJapproximate}
\end{equation}
which is the familiar weak-tunneling formula expected at `large' charging energy.  (See Ref.~\onlinecite{SchoenTunnelJunctionsReview} for a derivation in a junction with many weakly coupled channels.)  It is therefore reasonable to take $H = H_C + H_J$ with Eqs.~\eqref{HC} and \eqref{eq:hj-gen} for both the Josephson- and charging-energy-dominated regimes of interest here.

Importantly, the Hilbert space factorizes into even and odd fermion-parity sectors that do not mix under the dynamics of $H$.  The space within each sector can be spanned with either number $\hat n$ eigenstates, which we denote by $|n\in {\rm even}\rangle$ and $|n \in {\rm odd}\rangle$, or phase $\hat \varphi$ eigenstates with even and odd parity, denoted $|\varphi,e\rangle$ and $|\varphi,o\rangle$.  These bases are related by \cite{MajoranaTransportWithInteractions1}
\begin{equation}
  | \varphi, e \rangle  \propto  \sum_{n \text{ even}} e^{- i \varphi n / 2} |
  n \rangle,~~~~
  | \varphi, o \rangle  \propto  \sum_{n \text{ odd}} e^{- i \varphi n / 2} |
  n \rangle .
\end{equation}
We thus see that in the phase basis, fermion parity is reflected in how the wavefunctions behave under winding the eigenvalue $\varphi$ by $2\pi$; even states are invariant while odd states acquire a minus sign \cite{MajoranaTransportWithInteractions1}. Note also that the Hamiltonian  distinguishes opposite-parity states only through charging energy---this is crucial for capturing topological superconductivity within our phenomenological model.

In practice, the junction is expected to support a few channels, each with $T_{i}$'s that can be tuned by voltage on a local gate (valve) acting in the non-proximitized nanowire segment in Fig.~\ref{SetupFig} \cite{Gatemon,LangeNanowireJosephson}.  We refer to the valve as `closed' if the gate is adjusted so that the barrier is depleted of carriers, creating an SIS junction with all $T_i \approx 0$.  Charging energy $E_C$ for the essentially isolated mesoscopic island then dominates over the Josephson energy $E_J$.  Conversely, an `open' valve restores carriers to the barrier and hence boosts the $T_{i}$'s, ideally with at least one of the channels approaching unity transmission.  We assume that the maximal Josephson energy in this regime greatly exceeds the charging energy, i.e., $E_J \gg E_C$.

It is instructive to first examine the physics of an open valve in the limit $E_J/E_C \rightarrow \infty$.  In this case, the phase difference between island and bulk superconductor, $\hat \varphi$, becomes a classical variable locked at zero.  The island then forms a topological superconductor with Majorana zero modes $\gamma_{1,2}$---which satisfy $\gamma_i = \gamma_i^\dagger$ and $\{\gamma_i,\gamma_j\} = 2\delta_{ij}$---localized to its ends as Fig.~\ref{SetupFig}(a) sketches.  (We neglect the exponentially small overlap among these modes for now, though such corrections will become important in later sections.)  Accordingly, our phenomenological Hamiltonian supports degenerate, opposite-parity ground states $|\varphi = 0,e\rangle$ and $|\varphi = 0,o\rangle$ characteristic of this phase.
The Majorana zero-mode operators toggle the system between these ground states: $\gamma_i  |\varphi = 0,e/o\rangle \propto |\varphi = 0,o/e\rangle$.  

At large but finite $E_J/E_C$, residual charging energy weakly splits this degeneracy (see Appendix~\ref{sec:junction-compare}).  Adopting the form of $H_J$ in Eq.~\eqref{HJapproximate} yields a splitting \cite{Koch}
\begin{equation}
  \Delta E \approx \frac{32}{(2\pi^2)^{1/4}}E_C \left(\frac{E_J}{E_C}\right)^{3/4} e^{-\sqrt{8 E_J/E_C}}\cos(\pi n_0).
  \label{ECsplitting}
\end{equation}
Notice that $\Delta E$ vanishes \emph{exponentially} as $E_J/E_C$ increases so that the topological degeneracy remains robust provided $E_J/E_C \gg 1$ \cite{TopTransmon,BraidingWithoutTransport,BeenakkerBraiding,Hutzen}.  Figure~\ref{EnergyLevels}(a) sketches the low-lying energy levels at $E_J/E_C \gg 1$ versus the back-gate voltage $V_g$ that tunes $n_0$.  Red and blue curves respectively correspond to even- and odd-parity states.  It is worth emphasizing that Eq.~\eqref{ECsplitting} likely overestimates the topological degeneracy splitting in these setups.  A more efficient suppression of $\Delta E$ appears if one employs the more general form of the Josephson Hamiltonian in Eq.~\eqref{eq:hj-gen} with all $T_i \rightarrow 1$, though the results are otherwise qualitatively similar; see Appendix \ref{sec:junction-compare}.  Our phenomenological model also assumes that the two ends of the junction couple weakly on the scale of the pairing gap since only Cooper-pair tunneling is included.  The validity of this assumption is not obvious in the maximally open valve configuration.  The opposite strong hybridization limit requires a more microscopic treatment of the junction that will not be pursued here but may suppress the degeneracy splitting further still.  Finally, we note that the parameter $E_C$ may renormalize downward as the $T_i$'s increase from zero, thus also reducing the splittings compared to estimates based on $E_C$ extracted for an isolated island (see also Sec.~\ref{OptimizationStrategies} below).

\begin{figure*}
\includegraphics[width=0.9\textwidth]{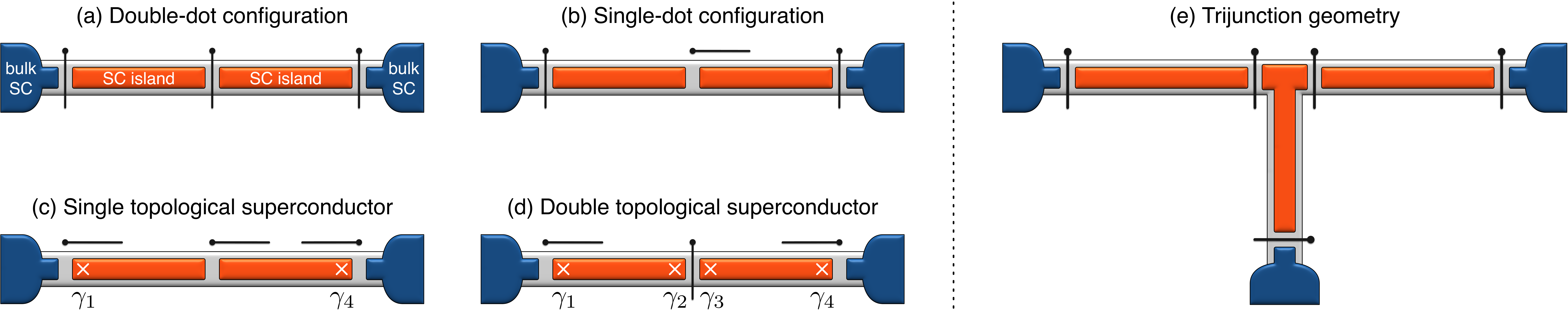}
\caption{Geometries utilized for the proposed milestone experiments.  (a)-(d) Two-island setup with gate-tunable valves that electrically tune the Josephson-to-charging energy ratios for each pair of adjacent superconducting regions.  Various interesting regimes are possible depending on the valve configurations: When cut off from the outer bulk superconductors the islands can (a) form independent double quantum dots or (b) effectively combine into a single dot.  Restoring the connection to the outer superconductors quenches charging energy and generates topological superconductivity on the islands.  Either two or four Majorana zero modes (denoted by $\times$'s) appear depending on the middle-valve configuration as shown in (c) and (d).  (e)~Trijunction geometry that enables braiding through similar valve manipulations.}
\label{TwoIslandSetupFig}
\end{figure*}

As the valve is closed more and more, the ratio $E_J/E_C$ decreases to values of order one, leading to the situation in Fig.~\ref{EnergyLevels}(b).  Going further, the fully closed valve gives $E_J/E_C \rightarrow 0$.  Energy eigenstates in this limit are characterized by well-defined island charges as illustrated in  Fig.~\ref{EnergyLevels}(c).  We will \emph{not} refer to the island as topological in this charging-energy-dominated regime.  Except with fine-tuning, the system admits a unique ground state, and hence Majorana zero modes are absent \footnote{To be precise, we define `Majorana zero modes' to be localized operators that guarantee protected ground-state degeneracy, up to exponentially small corrections, in the full Hamiltonian (including interactions).}.  The lack of an offset for even- versus odd-charge parabolas in Fig.~\ref{EnergyLevels}(c) does, however, reflect a remnant of the zero modes that exist when charging energy is quenched; indeed the island can still accommodate an odd number of electrons without paying the pairing energy $\Delta$ \cite{MajoranaTransportWithInteractions1, AlbrechtExponential}.  This property alters the even-odd pattern for Coulomb blockade relative to a more conventional superconducting dot.  Small residual Josephson coupling $E_J/E_C \neq 0$ merely produces avoided crossings between levels with common charge parity.

Figures~\ref{EnergyLevels}(a-c) illustrate an essential feature in our proposal: The topologically degenerate parity eigenstates smoothly and uniquely evolve into non-degenerate charge states upon closing an initially open valve, and vice versa. This is the phenomenon of parity-to-charge conversion introduced briefly in Sec.~\ref{Introduction}.  As a trial application, suppose that one begins in a closed-valve configuration with the isolated island relaxed into its unique charge ground state.  Slowly opening the valve nucleates Majorana zero modes and deterministically (and reproducibly) initializes the system into one of the two degenerate parity ground states.  Passing instead from degenerate parity states to charge states in a useful way requires an important implicit assumption invoked throughout this paper: When a valve closes, the energy difference between the lowest two charge states must not exceed the quasiparticle gap $\Delta_{\rm bulk}$ for the adjacent bulk superconductor, which prevents charge from  escaping from the islands.  Such a condition is always satisfied if $\Delta_{\rm bulk} > E_C$.  Figure~\ref{EnergyLevels}(c) illustrates this criterion graphically.

\subsection{Two-island geometry}

We now extend the preceding discussion to the two-island geometry shown in Figs.~\ref{TwoIslandSetupFig}(a-d). The device contains three valves to control the Josephson coupling to the leads and between the two islands.

We model the two-island setup with a Hamiltonian $H' = H_C' + H_J'$ analogous to that used earlier.  Let $\hat n_{L/R}$ denote the electron number operator for the left/right island, and $e^{i \hat \varphi_{L/R}}$ move a Cooper pair from the left/right island into the adjacent outer bulk superconductor.  Coulomb interactions are described by
\begin{equation}
  H_C' = \sum_{a = L/R} E_{C} (\hat n_a - n_{0,a})^2, 
\end{equation}
where $n_{0,L/R}$ are tunable offset charges for the islands. We neglect cross-capacitance, which is small in this geometry. In practice, we find experimentally that cross capacitances in a double-dot nanowire geometry are typically less than 0.1 of the total capacitance for each dot.

 For the Josephson coupling we write
\begin{equation}
  H_J' = H_{J,L} + H_{J,R} + H_{J,\rm inter}.
\end{equation}
The first two terms represent straightforward generalizations of Eq.~\eqref{eq:hj-gen} for the left and right junctions while $H_{J,\rm inter}$ hybridizes the two islands.  This last term requires some additional care.  In the Josephson-dominated regime the islands form topological superconductors that couple not only through Cooper-pair tunneling, but also through the Majorana-mediated fractional Josephson effect that coherently transfers unpaired electrons yielding an anomalous $4\pi$-periodic current-phase relation \cite{1DwiresKitaev}.  We therefore model inter-island Josephson coupling via
\begin{eqnarray}
  H_{J,\rm inter} &=&   
  -\Delta \sum_{i = 1}^N \sqrt{1 - T_{i,M} \sin^2 \left[(\hat\varphi_R-\hat\varphi_L) / 2 \right]}
  \nonumber \\
  &+&  H_{\rm FJE}. 
  \label{eq:ht}
\end{eqnarray}
In the first line $T_{i,M}$ denotes the transmission probabilities for the barrier separating the islands; the second line denotes the fractional Josephson coupling.  If all transmission probabilities are small the latter reduces to the familiar expression \cite{1DwiresKitaev}
\begin{eqnarray}
  H_{\rm FJE} &=&   - \Gamma \sigma^x \cos \left[(\hat\varphi_R-\hat\varphi_L) / 2 \right],
  \label{eq:hm}
\end{eqnarray}
where $\Gamma >0$ and the operator $\sigma^x$ flips the island parities when acting on phase eigenstates.  (Away from the weak-transmission limit higher harmonics can arise, similar to the usual Josephson coupling.)  The central valve in Figs.~\ref{TwoIslandSetupFig}(a)--(d) turns both $T_{i,M}$ and $\Gamma$ `on' and `off' in tandem.

Adjusting the three valves allows access to several useful configurations.  With all three valves closed [Fig.~\ref{TwoIslandSetupFig}(a)] the islands form  independent quantum dots (except for cross capacitance) with energy eigenstates characterized by well-defined charges on each island.  Opening the central valve [Fig.~\ref{TwoIslandSetupFig}(b)] strongly Josephson-couples the islands, which then behave as a single isolated dot with energy eigenstates carrying fixed total charge.  Opening the outer valves [Fig.~\ref{TwoIslandSetupFig}(c)] converts the isolated dot into a topological superconductor hosting a pair of Majorana zero modes that encode degeneracy between even- and odd-parity ground states.  And finally, re-closing the middle valve [Fig.~\ref{TwoIslandSetupFig}(d)] cuts the topological superconductor in half, nucleating a second pair of Majorana zero modes that yield a two-fold degeneracy \emph{within} each total parity sector.  The flexibility afforded by this two-island geometry enables both fusion-rule detection and topological qubit validation.  Moreover, the manipulations possible in this setting generalize naturally to more elaborate geometries including the trijunction in Fig.~\ref{TwoIslandSetupFig}(e) employed for braiding.

\subsection{Readout methods}
\label{ReadoutSec}

Detecting the state formed by a topological superconductor after some prescribed sequence of manipulations can be achieved by closing the appropriate  valves to convert degenerate parity ground states segment into charge states (Figs.~\ref{SetupFig}, \ref{EnergyLevels}), which can then be readily measured via charge detection \cite{vanderWiel,Hanson}.

The most direct readout method uses charge sensing of an isolated island, accomplished with capacitive coupling to a proximal sensor such as a quantum point contact or a quantum dot. This technique is
routinely used for readout in spin qubits \cite{Barthel,Medford,Eng}. Quantum-dot charge sensors have also been implemented in nanowires, obtaining couplings similar to the GaAs case \cite{Hu1,Hu2}. Based on these experiments, high-quality charge readout [signal-to-noise ratio $\sim 5$, corresponding to 99\% fidelity] should be achievable with $\tau_M = 1~\mu$s integration time.  Single-shot readout is possible when the charge state's lifetime---which is set by poisoning events and relaxation processes---exceeds $\tau_{M}$ at the measurement point.  Even if this criterion is not satisfied, readout may be performed over several identical cycles to obtain the expectation value for the island charge.

Alternatively, one can employ dispersive readout, which measures \emph{changes} in the average charge $\langle Q \rangle$ of a device by probing its quantum capacitance $\partial \langle Q \rangle / \partial V_g$.  This technique is experimentally well established both for semiconductor quantum dots~\cite{PeterssonDispersive,DispersiveReadoutQD}, including InAs nanowires~\cite{JungDispersive}, as well as conventional Cooper-pair boxes~\cite{DispersiveReadoutCPB}. To adapt dispersive readout for parity measurements, one could operate the superconducting island in the intermediate regime $E_J/E_C \sim 1$ and tune close to an avoided crossing for one of the parity states [see Fig.~\ref{EnergyLevels}(b)]. At such a point, the quantum capacitance is maximal for the parity associated with the avoided crossing, while the lower-energy, opposite-parity state exhibits a stable charge configuration and thus admits a rather small quantum capacitance.  Readout can therefore proceed by detecting the capacitance difference between
these states \footnote{We note that dispersive readout differs from the parity readout suggested for transmon \cite{TopTransmon} or flux qubits \cite{Hassler} at large Josephson energies $E_J/E_C$, in which a parity-dependent modification of the transition frequency between the energy levels would be detected [see Fig.~\ref{EnergyLevels}(a)].}. Based on experiments in InAs nanowires, dispersive parity readout should be possible with less than a millisecond of integration time using standard amplifiers \cite{JungDispersive}, or much faster with nearly quantum-limited amplifiers \cite{StehlikDispersive}.

Finally, we consider an approach based on continuous cyclic gate operation, which requires fast gate operation but allows slow, time-averaged readout.  For now we simply provide a glimpse of the recipe; a detailed discussion appears in Sec.~\ref{Pump}.  Consider again the double-dot configuration in Fig.~\ref{TwoIslandSetupFig}(a), and suppose that we wish to distinguish two different charge states $C_1$ and $C_2$ that evolve from topologically degenerate states under parity-to-charge conversion.  Under an appropriate gate-voltage cycle, one can arrange to have charge $q$ shuttled across the system if $C_1$ appears, but not if $C_2$ appears.  Repeating the entire process with a frequency $f$ then produces a pumping current $I = P q f$, where $P$ denotes the probability for obtaining state $C_1$.  Majorana modes enter the story in that they fix very precise values for the probability $P$ dependent on how the user manipulated the topological superconductors prior to readout.  One can thereby detect the ensemble-averaged outcome of fusion-rule experiments, topological qubit manipulations, and even braiding through dc current measurements.

\subsection{Operating requirements and optimization strategies}
\label{OptimizationStrategies}

Varying Josephson-to-charging energy ratios via gate-tunable valves provides an all-electrical alternative to the related proposals in Refs.~\onlinecite{TopTransmon,BraidingWithoutTransport,BeenakkerBraiding}.  The latter studies operate solely in the $E_J \gg E_C$ regime, control $E_J/E_C$ using local magnetic-flux tuning in split junctions, and perform readout by measuring the frequency of a transmission line resonator.  Our approach, by contrast, utilizes charge-based readout methods outlined in Sec.~\ref{ReadoutSec} and requires tuning between the Josephson- and charging-energy-dominated limits.  More precisely, we demand 
\begin{equation}
  \left(\frac{E_J}{E_C}\right)_{\rm max} \gg 1,~~~\left(\frac{E_J}{E_C}\right)_{\rm min} \lesssim 1,~~~E_C \gtrsim k_B T,
  \label{ValveRequirements}
\end{equation}
where $E_J$ refers to the Josephson coupling between an island and a trivial bulk superconductor \footnote{Tunability of inter-island Josephson couplings faces less stringent requirements, at least in the devices that we will study in this paper.  For instance, in Fig.~\ref{TwoIslandSetupFig}(c) inter-island Josephson coupling merely joins two topological islands whose charging energy is already quenched and thus need not greatly exceed $E_C$ when the intervening valve is open.}.
The first criterion in Eq.~\eqref{ValveRequirements} minimizes unwanted topological degeneracy splitting due to residual Coulomb effects in open-valve configurations (see Sec.~\ref{SingleIslandSec}) and is essential for providing a reasonable window between the upper and lower speed limits at which our proposed protocols should proceed; see Sec.~\ref{RealisticCase} and Ref.~\onlinecite{Hell} for in-depth discussions.  The second and third criteria facilitate readout of isolated islands bordered by closed valves, and for these criteria $E_C$ refers to closed-valve, unrenormalized charging energies (see below) \footnote{When discussing readout, e.g., in the bottom steps of Fig.~\ref{FusionRuleFig}, for conceptual clarity we typically appeal to the regime $\left(E_J/E_C\right)_{\rm min} \ll 1$ corresponding to (experimentally accessible) fully closed valves.  While fully quenching Josephson energy at these stages facilitates charge sensing, this extreme limit is not necessary for all readout options (notably dispersive readout); see Sec.~\ref{ReadoutSec}.}.  The extent to which one can satisfy these inequalities constitutes an important figure of merit for our scheme.

In practice, we expect the second and third inequalities to be readily satisfied:  One can efficiently reduce $E_J$ near zero by fully depleting a given barrier, while charging energies for isolated islands can easily exceed temperature (typical bare $E_C$ values correspond to a few tenths of a Kelvin).  The first inequality, on the other hand, requires more care in our setups, though we anticipate that values of $(E_J/E_C)_{\rm max} \sim 10-100$ can be realized with fully open barriers and optimized capacitance.

First off, as alluded to in Sec.~\ref{SingleIslandSec}, $E_C$ is expected to renormalize downwards as a valve is opened.  To contextualize this point, for a superconducting island coupled to a \emph{normal} lead via a single fully transmitting channel, the charging energy is known to renormalize all the way to zero \cite{Feigelman_ECrenorm}.  For the present SNS junctions of interest, the extent of this renormalization is still unknown, but it is not unreasonable to expect substantial boosts in the achievable $E_J/E_C$ ratios from this effect alone.  

\begin{figure}[b]
\includegraphics[width=\columnwidth]{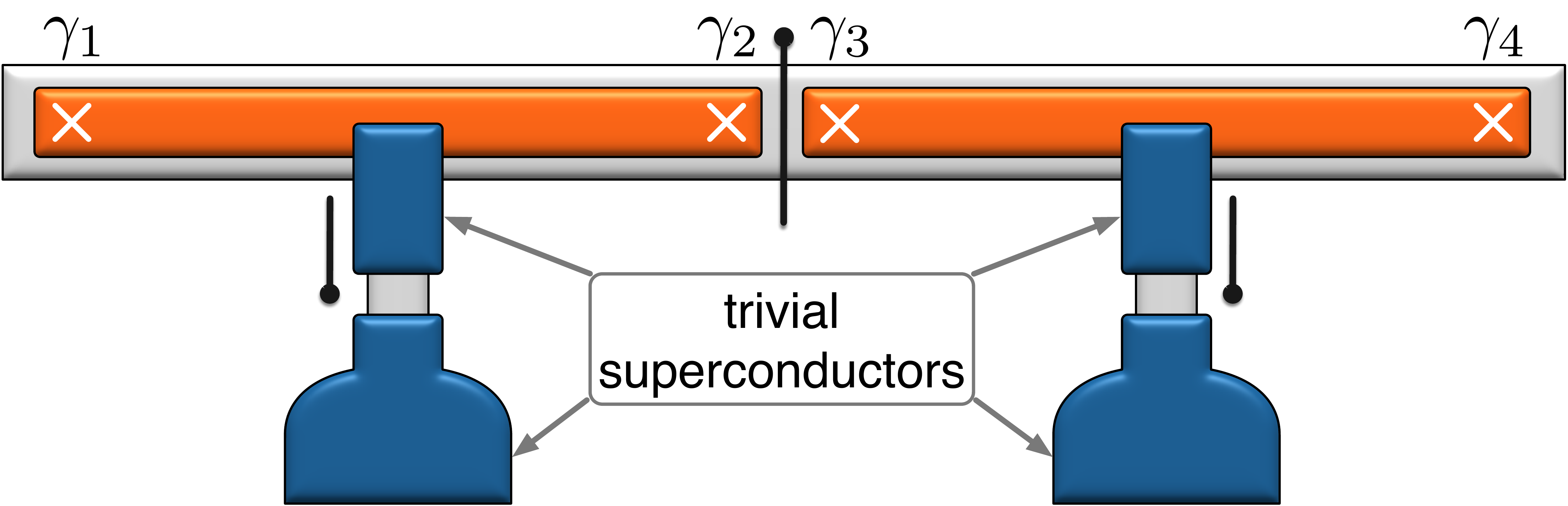}
\caption{Alternative two-island setup involving trivial superconducting wires forming gate-tunable Josephson junctions with conventional bulk superconductors.  This shunt-junction configuration is expected to more efficiently quench the island charging energy in the open-valve configuration shown here.  Indeed the vertical wires can support $N\gg 1$ channels (unlike the horizontal wire hosting Majorana modes), thus enhancing the maximum Josephson energy $E_J$.}
\label{ShuntFig}
\end{figure}

Feasible options are available, however, even in the worst-case scenario of negligible renormalization.  Most simply, one can increase the length of the islands and/or thickness of the superconducting shells to increase their capacitance and decrease $E_C$ \footnote{The capacitance $C$ for a wire of length $L$ and radius $a$ is proportional to $L/\log(L/a)$.  For typical nanowires with $L\sim1~\mu\mathrm{m}$ and $a\sim100~\mathrm{nm}$, doubling the length of a wire thus decreases $E_C\sim1/C$ by a factor of approximately 1.5}.  One can also connect the shells to larger superconducting islands to efficiently reduce $E_C$; Refs.~\onlinecite{Gatemon,LangeNanowireJosephson} employed this strategy in closely related non-topological transmon-like qubits with reported charging energies of a few mK.  [Contrary to those works, however, we require the third inequality in Eq.~\eqref{ValveRequirements}, so $E_C$ should not be decreased too substantially.]  Yet another option is to connect the islands to \emph{trivial} superconducting wires that form gate-tunable Josephson junctions with conventional grounded bulk superconductors, as illustrated in Fig.~\ref{ShuntFig} for a two-island setup.  Because these `shunt junctions' are trivial on both ends, the wire mediating the Josephson coupling may host many channels and yield correspondingly large $E_J$ values without spoiling the Majorana physics of interest.  By contrast, the upper wire in Fig.~\ref{ShuntFig} that harbors the Majorana modes should support only a few channels and thus carry a relatively large normal-state resistance.  

While we hereafter concentrate on minimalist setups based on Figs.~\ref{SetupFig} and \ref{TwoIslandSetupFig}, the above variations offer practical alternatives that facilitate satisfying the operating requirements of Eq.~\eqref{ValveRequirements} and may be straightforwardly substituted into all of our proposed experiments.


\section{Fusion-rule detection}
\label{FusionRuleSec}

\subsection{Motivation}

One of the signature properties of non-Abelian anyons is their exotic exchange statistics: Braiding these particles rotates the host system's quantum state within a degenerate ground-state manifold.  An intimately related and equally fundamental property is that non-Abelian anyons exhibit nontrivial `fusion rules'.  That is, they can coalesce to yield multiple quasiparticle types. One can understand the connection between non-Abelian statistics and nontrivial fusion rules by examining the evolution of ground states.  Non-Abelian braiding properties require that the anyons possess zero-energy degrees of freedom that \emph{non-locally} encode ground-state degeneracy.  Bringing two non-Abelian anyons together hybridizes these degrees of freedom and thus generically splits the initial ground-state manifold into a (smaller) ground-state set and excited states.  Each distinct energy then corresponds to a different possible fusion product for the anyons.

\begin{figure}
\includegraphics[width=3in]{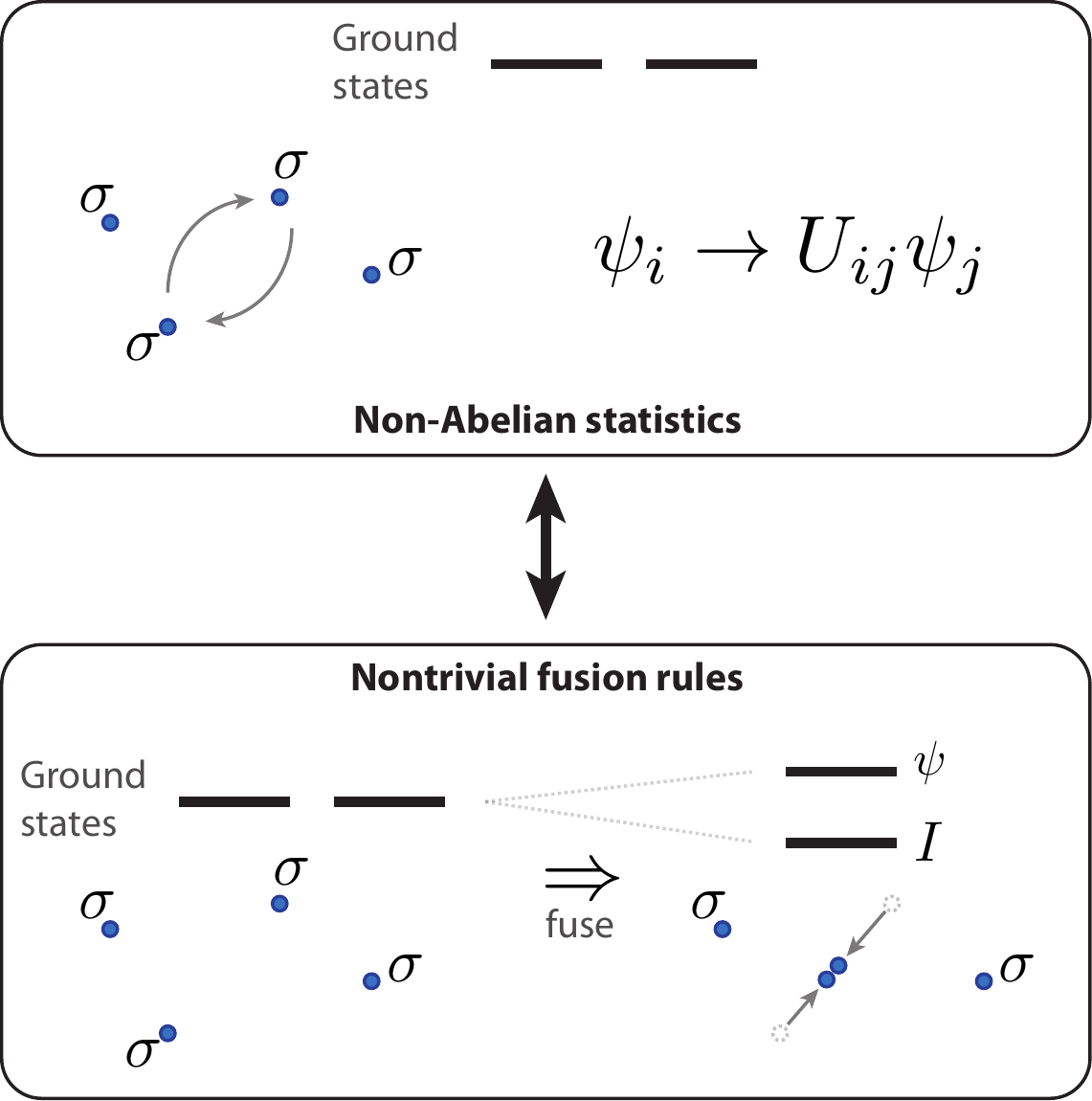}
\caption{Correspondence between non-Abelian statistics and nontrivial fusion rules for Ising anyons $\sigma$ that obey $\sigma \times \sigma = I+\psi$.  Top: Each of the four Ising anyons shown binds a Majorana zero mode.  With fixed total parity the system then admits two degenerate ground states $\psi_{i = 1,2}$.  Adiabatically braiding a pair of anyons sends $\psi_i \rightarrow U_{ij}\psi_j$, where $U_{ij}$ denotes the braid matrix.  Bottom: Alternatively fusing the pair of anyons hybridizes the associated Majorana modes and splits the twofold ground-state degeneracy.  The lower and upper states correspond to the Ising anyons coalescing into the identity $(I)$ and fermion $(\psi)$ fusion channels.  While fusion and braiding are intimately linked, establishing the former experimentally is expected to be much simpler; see Figs.~\ref{FusionRuleFig} and \ref{PumpFig}.
}
\label{FusionRuleStatisticsFig}
\end{figure}

In the present context the ends of 1D topological superconductors behave as Ising non-Abelian anyons, denoted $\sigma$, and the Majorana zero modes that they bind constitute the degrees of freedom guaranteeing degeneracy.  These anyons obey the fusion rule
\begin{equation}
  \sigma \times \sigma = I + \psi,
  \label{fusionrule}
\end{equation}
which indicates that two $\sigma$'s can fuse into the trivial particle $I$ (i.e., they can annihilate) or a fermion $\psi$ \footnote{For completeness we note that an Ising anyon theory possesses other fusion rules, albeit less interesting for our purposes here.  These include $\psi \times \psi = I$ and $\sigma \times \psi = \sigma$. }.  The fusion outcomes on the right side simply reflect the fact that two Majorana modes brought together form an ordinary, finite-energy fermionic level that can be either vacant ($I$) or filled ($\psi$).  Figure~\ref{FusionRuleStatisticsFig} summarizes the relation between statistics and fusion rules for this case.

A comparison may help contextualize the connection between non-Abelian statistics and fusion. Consider a pair of isolated spin $s = 1/2$ moments each with degenerate `up' and `down' levels.  When brought together the spins experience exchange coupling that splits the degeneracy into (say) a lower singlet with total spin $s_{\rm tot} = 0$ and an upper triplet with $s_{\rm tot} = 1$.  These quantum numbers follow from the familiar angular momentum addition rule $1/2\times 1/2 = 0 + 1$, which appears reminiscent of Eq.~\eqref{fusionrule}.  A crucial distinction exists, however. In the spin example, a perturbation such as a local Zeeman field splits the degeneracy even when the spins are well separated. By contrast, \emph{no} local perturbation can split the degeneracy encoded by distant non-Abelian anyons since the corresponding ground states are locally indistinguishable.  Fusion of non-Abelian excitations thus involves degeneracy lifting of a very special nature.

Fusion rules like Eq.~\eqref{fusionrule} can even serve to \emph{define} the non-Abelian character of anyons. Specifically, non-Abelian statistics implies multiple fusion outcomes and vice versa \cite{SymmetryDefects,Rowell}.  References \onlinecite{AliceaBraiding,Ruhman} proposed schemes for detecting fusion rules in nanowires and cold atoms setups.  Related ideas in a quantum-Hall context appear even earlier \cite{Bishara}.  Below we develop two new approaches for demonstrating this milestone based on the two-island setup from Figs.~\ref{TwoIslandSetupFig}(a)-(d).  

\subsection{Fusion rules via charge sensing}
\label{FusionRuleChargeSensing}

\subsubsection{Idealized limit}
\label{IdealLimit}

We first outline the charge-sensing fusion-rule experiment in the ideal case, i.e., with exactly zero overlap between Majorana zero modes (when present), infinite tunability of Josephson-to-charging energy ratios, no quasiparticle poisoning, and limitless patience of experimentalists conducting the measurements.  Under these assumptions all manipulations described below proceed purely adiabatically.  Section~\ref{RealisticCase} discusses corrections resulting from inevitable non-idealities.

The protocol begins from the topmost configuration of Fig.~\ref{FusionRuleFig}, wherein the central valve remains open but the outer valves are closed.  In this configuration the two islands---which strongly Josephson-couple to one another---effectively form a single Coulomb-blockaded quantum dot with a unique ground state $|Q_{\rm tot}\rangle$ characterized by a fixed total charge $Q_{\rm tot}$.  (Here and below we assume that potentials on the islands are not fine-tuned to give accidentally degenerate charge states.)  Suppose that we initialize the system into this ground state and, as a warm-up control experiment, carry out the following steps depicted in the right path of the figure:

\begin{figure}
\includegraphics[width=\columnwidth]{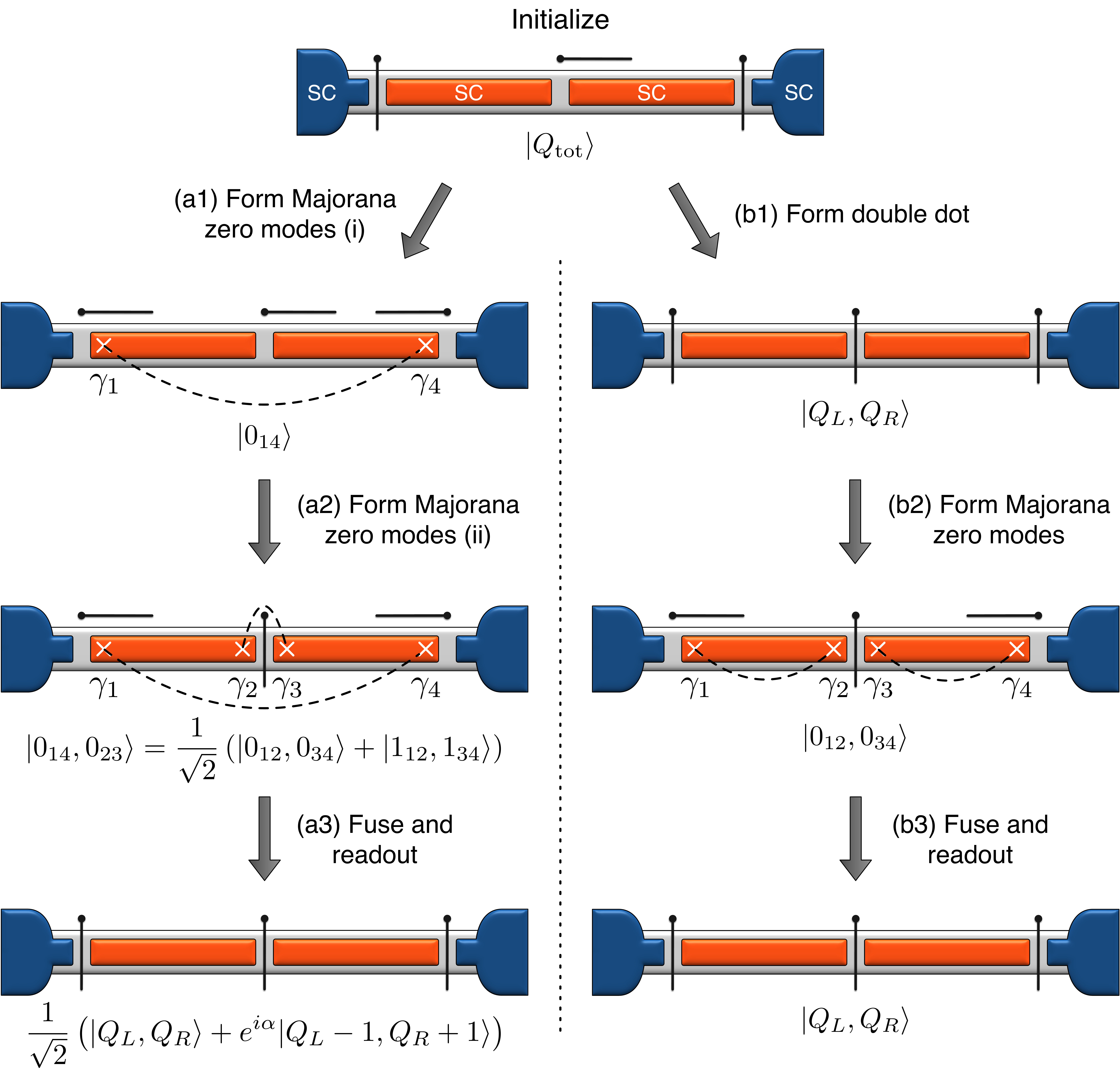}
\caption{Protocol to detect the Ising-anyon fusion rule $\sigma \times \sigma = I + \psi$ via charge sensing.  The device contains two superconducting islands (red) Josephson coupled to outer trivial bulk superconductors (blue); intervening gate-tunable valves control the Josephson-to-charging energy ratios.  The system is initialized into a unique ground state by opening the middle valve while closing the outer valves.  In the control experiment (right path), the middle valve closes \emph{before} opening the outer valves, thus nucleating out of the vacuum two pairs of Ising anyons binding Majorana modes $\gamma_{1,2}$ and $\gamma_{3,4}$ as indicated by dashed lines.  Re-closing the outer valves fuses these same pairs back into the vacuum, corresponding to the $I$ fusion channel.  Charge readout on the islands is correspondingly deterministic.  Instead closing the middle valve \emph{after} opening the outer valves (left path) first nucleates Ising anyons binding $\gamma_{1,4}$ and later $\gamma_{2,3}$.  Closing the outer valves now fuses the anyons in a nontrivial way that accesses both the $I$ and $\psi$ fusion channels.  This leads to an equal-amplitude superposition of states in the measurement basis and thus probabilistic charge readout.  }
\label{FusionRuleFig}
\end{figure}

{\bf (b1) Close the central valve.} The system then evolves into a double dot with unique ground state $|Q_L,Q_R\rangle$, where $Q_{L/R}$ denote the charges on the left/right islands (which of course sum to $Q_{\rm tot}$).

{\bf (b2) Open the outer valves.}  Quenching the charging energy in this manner drives the islands into topological superconductors with Majorana zero modes $\gamma_{1,\ldots,4}$ enumerated from left to right as in the figure.  Let
\begin{equation}
  f_{12} = (\gamma_1+i\gamma_2)/2,~~~~f_{34} = (\gamma_3+i\gamma_4)/2
  \label{fermions}
\end{equation}
denote complex fermion operators built from these Majorana modes, and label the corresponding occupation numbers by $n_{12}$, $n_{34}$.  Since the topological superconductors emerged from decoupled islands, it is conceptually useful to envision this step as pulling Ising anyons binding $\gamma_{1}$ and $\gamma_2$ out of the vacuum, and similarly for $\gamma_{3}$ and $\gamma_4$.   (Figure~\ref{FusionRuleFig} links such Majorana pairs with dashed lines.)
The system therefore evolves into a state with well-defined values of $n_{12}$ and $n_{34}$.  Without loss of generality we assume that these occupation numbers both vanish, yielding a wavefunction $|0_{12},0_{34}\rangle$ \footnote{The precise \emph{initial} values of the occupation numbers are convention-dependent and thus not especially meaningful.  We simply set them to zero here and below to streamline the presentation.}.  We have now prepared a system with four non-Abelian anyons and initialized into one of the two degenerate ground states available when total fermion parity is fixed.  

{\bf (b3) Close the outer valves.}  This final step removes the ground-state degeneracy by resurrecting charging energy---thus fusing the very same pairs of non-Abelian anyons nucleated from the vacuum above.  The system then simply retraces its footsteps back to the double-dot state $|Q_L,Q_R\rangle$ initialized at the end of step (b1).  A charge-sensing measurement of either island here clearly yields a unique outcome.  In terms of the fusion rule in
Eq.~\eqref{fusionrule}, the sequence concluded here deterministically reveals only fusion channel $I$.  Of course the same deterministic charge measurement would \emph{also} occur for non-topological superconducting islands.

This so-far unremarkable conclusion establishes a very useful baseline.
A much more interesting scenario appears if we merely swap the order in which we apply steps (b1) and (b2) from the preceding control protocol, thereby following the left path in Fig.~\ref{FusionRuleFig}.  Let us begin anew from the state $|Q_{\rm tot}\rangle$ in the topmost configuration of the figure and explore this alternative sequence in some detail:

{\bf (a1) Open the outer valves.}  Because the islands are now well Josephson coupled, this process generates a \emph{single} topological superconductor and pulls Ising anyons binding Majorana zero modes $\gamma_1$ and $\gamma_4$ out of the vacuum.  One can assemble these modes into a complex fermion operator
\begin{equation}
  f_{14} = (\gamma_1 + i\gamma_4)/2
  \label{f14}
\end{equation}
whose occupation number $n_{14}$ takes on a unique value due to global fermion parity conservation.  The system therefore evolves into the quantum state $|0_{14}\rangle$.

{\bf (a2) Close the central valve.}  Quenching the Josephson coupling between the islands effectively slices the topological superconductor in half.  Consequently, a second pair of Ising anyons with Majorana zero modes $\gamma_2$ and $\gamma_3$ appears opposite the middle valve.  Similarly defining a fermion
\begin{equation}
  f_{23} = (\gamma_2+i\gamma_3)/2
  \label{f23}
\end{equation}
with occupation number $n_{23}$, we then arrive at the state $|0_{14},0_{23}\rangle$.  In the basis introduced in Eq.~\eqref{fermions} this state equivalently reads \footnote{To be precise we define here $|1_{12},1_{34}{\rangle} \equiv f_{12}^\dagger f_{34}^\dagger |0_{12},0_{34}{\rangle}$.}
\begin{equation}
  |0_{14},0_{23}\rangle = \frac{1}{\sqrt{2}}(|0_{12},0_{34}\rangle + |1_{12},1_{34}\rangle).
  \label{EntangledState}
\end{equation}
The superposition on the right side merely reflects a basis change related to so-called $F$-symbols for Ising anyons; see Appendix \ref{AnyonAppendix}.  In the ideal case where the topological ground-state degeneracy is exact and immune to noise, this state does not dephase.  We have prepared four non-Abelian anyons as before---in fact in an ostensibly identical setup---but, interestingly, initialized the system into a different degenerate ground state that maximally entangles the two decoupled islands.

{\bf (a3) Close the outer valves.}  Restoring charging energy once again removes the ground-state degeneracy; this process fuses the anyons binding $\gamma_1$ and $\gamma_2$, and also the pair binding $\gamma_3$ and $\gamma_4$.  Notably, these are \emph{not} the same pairs nucleated from the vacuum, contrary to the left sequence in Fig.~\ref{FusionRuleFig}, which allows us to now access both fusion channels on the right side of Eq.~\eqref{fusionrule}.  As we restore charging energy, $|0_{12},0_{34}\rangle$ evolves into the unique ground state $|Q_L,Q_R\rangle$, precisely as noted earlier.  The wavefunction $|1_{12},1_{34}\rangle$ instead evolves into an excited state with opposite charge parities on the islands compared to $|Q_L,Q_R\rangle$.  We will assume for concreteness that the gates are adjusted such that the excited state corresponds to a charge configuration $|Q_L-1, Q_R+1\rangle$.  With this assumption the state in Eq.~\eqref{EntangledState} becomes
\begin{equation}
  \frac{1}{\sqrt{2}}(|Q_L,Q_R\rangle + e^{i \alpha} |Q_L-1,Q_R+1\rangle),
  \label{superposition}
\end{equation}
where the first and second terms respectively reflect fusion into the $I$ and $\psi$ channels. Contrary to the state \eqref{EntangledState}, the superposition in \eqref{superposition} would dephase quickly in practice due to charge noise that results in the non-universal phase $\alpha$ indicated above.  This phase is irrelevant, however, since it is in any case invisible in our measurement scheme.  More importantly, charge measurement of an island at the end of this sequence is \emph{probabilistic}, returning even and odd electron numbers for a given island with equal probability.  The control experiment laid out earlier distinguishes this interesting scenario---which is deeply rooted in the nontrivial anyon fusion rules---from a boring `noisy' experiment with rapid charge fluctuations.  One may worry that the nontrivial protocol may happen to be more susceptible to charge fluctuations than the control protocol, since the intermediate steps are not exactly the same.  Additional control experiments can rule out this scenario.  For example, one can perform the protocols in Fig.~\ref{FusionRuleFig} with magnetic fields and/or back gate voltages adjusted so that the islands definitely do \emph{not} support Majorana modes; the measurement outcomes should then be identical for both the control and nontrivial sequences.  Observing this expected outcome would provide further evidence for fusion rules underlying the path dependence for the topological case.

It is worth remarking that probabilistic readout already suffices to demonstrate multiple fusion channels.  We stress that the \emph{equal} probabilities predicted here reflect more detailed topological information for the anyon theory.  In particular, these probabilities are given by ratios of appropriate quantum dimensions of particles in the topological quantum field theory or, equivalently in this case, by the absolute value of certain $F$-symbols \cite{TQCreview}.  We discuss these points in detail in Appendix \ref{AnyonAppendix}.

Although we are not probing non-Abelian statistics, there is a distinct `non-Abelian-ness' to this fusion rule experiment: The two sequences outlined above apply precisely the same adiabatic manipulations in a different order yet deliver dramatically different results.  In a reference non-topological system with no Majorana zero modes the ground state would generically be unique throughout, negating any such nontrivial path dependence under adiabatic evolution as nicely emphasized in Ref.~\onlinecite{Ruhman}.


\subsubsection{Practical considerations}
\label{RealisticCase}


With the ideal schemes fleshed out, we now highlight several practical issues pertinent for experimental implementation.  First, as with any experiment it is technically possible to obtain measurements in a non-topological setup that emulate our predictions based on nontrivial fusion rules.  We stress, however, that such a scenario is highly improbable and requires multiple levels of fine-tuning.  For one, accidental degeneracies would be required.  Suppose, for example, that a trivial system harbors a pair of fine-tuned zero-energy Andreev bound states located opposite the (closed) middle valve; for an illustration see Fig.~\ref{QubitFig}(b).  Although the right path of Fig.~\ref{FusionRuleFig} again trivially decouples the islands, the left path can entangle the two sides.  (Without degeneracy adiabatic evolution would preclude the latter.)  Importantly, however, the probabilities for even and odd charges at the end of the left path in Fig.~\ref{FusionRuleFig} would generically differ without \emph{additional} fine-tuning beyond that needed for accidental degeneracy.  In sharp contrast, the Ising-anyon fusion rules rigidly lock these probabilities together in the topological setup.  

Next we quantify the upper speed limit over which the steps in Fig.~\ref{FusionRuleFig} can be conducted while effectively remaining adiabatic.  Quite generally, this speed limit is set by the minimal gap for \emph{accessible} excited states (which should remain unoccupied) during a given stage of the protocol.  The rate for all steps must certainly fall below the pairing energy $\Delta$ for the islands to avoid generating unwanted Bogoliubov-de Gennes quasiparticles.  When closing valves in steps (a3), (b1), and (b3) from Fig.~\ref{FusionRuleFig}, we must also avoid producing spurious excited charge configurations that would corrupt initialization and readout of the anyons.  [For example, rapidly shutting the middle valve in (b1) would yield an entangled superposition of various charge states for the islands rather than the unique ground state $|Q_L,Q_R\rangle$.]  Charging energy $E_C$ determines the rate limit required to sidestep this type of corruption.  These considerations indicate that the protocols in Fig.~\ref{FusionRuleFig} approximate adiabatic evolution if carried out over a time scale
\begin{equation}
  t_{\rm protocol} \gg {\rm max}\left\{\frac{\hbar}{\Delta},\frac{\hbar}{E_C}\right\} ~~~~~\text{(adiabatic criterion)}.
  \label{adiabatic}
\end{equation}
This standard criterion can be justified by analyzing the protocol steps using the phenomenological model described in Sec.~\ref{GateSec}; see Ref.~\onlinecite{Hell}.
To quantify the times we assume a pairing gap near 1\,K and a charging energy on the same scale (which is not unreasonable for islands a few microns in length).  With these simple estimates Eq.~\eqref{adiabatic} becomes $t_{\rm protocol} \gg 10\, \text{ps}$.

In practice the fusion-rule protocols also face a \emph{lower} speed limit due to both quasiparticle poisoning and residual splitting of the ground-state degeneracies encoded by the Majorana modes.  To appreciate constraints from the former, suppose that one carried out the right control sequence from Fig.~\ref{FusionRuleFig} on a time scale much longer than the poisoning time $t_{\rm poisoning}$.  While the outer valves remain open stray quasiparticles can leak onto the islands, stochastically flipping the occupation numbers $n_{12}$ and $n_{34}$.  Distinction from the nontrivial right sequence from Fig.~\ref{FusionRuleFig} is then lost.  One must therefore maintain
\begin{equation}
  t_{\rm protocol} \ll t_{\rm poisoning} ~~~~~ \text{(poisoning limitation)}
  \label{poisoning}
\end{equation}
to suppress poisoning events that obscure fusion-rule detection.

The second issue---unintentional ground-state-degeneracy lifting---originates from several sources: $(i)$ wavefunction overlap of Majorana modes opposite a given island, $(ii)$ overlap of Majorana modes from neighboring islands, and $(iii)$ residual charging energy present even when the outer valves are maximally open.  Stages of the protocol where ideally exact degeneracies appear must transpire at a rate faster than the resulting energy splittings $E_{\rm splitting}$, i.e.,
\begin{equation}
  t_{\rm protocol} \ll t_{\rm splitting} ~~~~~ \text{(degeneracy limitation)}
  \label{degeneracy}
\end{equation}
with $t_{\rm splitting} \equiv \hbar/E_{\rm splitting}$.
Satisfying this inequality ensures that time evolution due to residual splitting is unimportant.

Let us quantify each of the three contributing sources noted above to get a rough sense for the time scales required:

$(i)$ Majorana overlap across a single island yields a splitting that scales, modulo an oscillatory prefactor, like $E_{\rm splitting}^{(i)} \sim \Delta  e^{-L/\xi}$ where $L$ and $\xi$ denote the size of the island and induced coherence length, respectively (see, e.g., supplementary information of Ref.~\onlinecite{SmokingGun}).  Taking $\xi = 500$~nm, $L = 3$~$\mu$m, and a 1 K pairing energy gives $E_{\rm splitting}^{(i)} \sim 3$~mK.  This yields a time scale $t_{\rm splitting}^{(i)} \sim 3$~ns.

$(ii)$ Overlap of the outermost Majorana modes $\gamma_1$ and $\gamma_4$ in Fig.~\ref{FusionRuleFig} generically contributes a tiny splitting compared to that considered above.  Splitting due to hybridization of $\gamma_2$ and $\gamma_3$ across the central (closed) valve is more important to estimate. Suppose that the nanowire hosts carriers with effective mass $m$, spin-orbit coupling $\alpha$, and a Zeeman field $V_{Z}$ that exceeds the proximity-induced pairing energy $\Delta$ as required for topological superconductivity in the islands \cite{1DwiresLutchyn,1DwiresOreg}.  We model the barrier between the islands as a region of length $W$ with a chemical potential shifted below the band bottom by an amount $U$ that is large compared to other energy scales for the wire.  The overlap of the inner Majorana modes then scales as $e^{-W/\xi_U}$ with $\xi_U = \hbar/\sqrt{2m U}$.  Moreover, using the canonical single-band wire Hamiltonian from Refs.~\onlinecite{1DwiresLutchyn,1DwiresOreg} in the large-field limit $V_{Z} \gg m \alpha^2,\Delta$ for simplicity, we find a splitting of \footnote{The wire Hamiltonian reduces to a toy model for spinless $p$-wave superconductor in the large-field limit.  We extract the splitting quoted in Eq.~\eqref{Eii} by solving the analogous problem for the latter using parameters obtained from the former.}
\begin{equation}
  E^{(ii)}_{\rm splitting} \approx \Delta \left(\frac{m \alpha \xi_U}{\hbar}\right)e^{-W/\xi_U}.
  \label{Eii}
\end{equation}
Consider reasonable parameters $\Delta = 1$~K, $W = 100$~nm, $U = 2$~meV, $\alpha = 2\times 10^4~{\rm m/s}$, and $m = 0.05 m_e$ ($m_e$ denotes the bare electron mass) as appropriate for wurtzite InAs wires \cite{WurtziteMass}.  These values yield $E^{(ii)}_{\rm splitting} \approx 1$ mK, corresponding to a time sale $t_{\rm splitting}^{(ii)} \approx 7$~ns.

$(iii)$ In the configuration with four Majorana modes---i.e., after steps (a2) and (b2) from Fig.~\ref{FusionRuleFig}---the two available ground states split by an amount given roughly by \cite{Koch}
\begin{equation}
  E^{(iii)}_{\rm splitting} \approx \frac{32}{(2\pi^2)^{1/4}}E_C \left(\frac{E_J}{E_C}\right)^{3/4} e^{-\sqrt{8 E_J/E_C}}
  \label{Eiii}
\end{equation}
per island due to residual charging energy $E_C$ that is not perfectly quenched by the Josephson coupling $E_J$ to the outer superconductors.  The latter may be inferred from measured critical currents via $I_c = 2e E_J/\hbar$.  
To get a feel for numbers, preliminary measurements suggest critical currents in the range $I_c \sim 100 - 200$~nA with corresponding Josephson energies $E_J \sim 2-5$~K.  Suppose that $E_C = 0.5$~K.  With $E_J = 2$~K one gets  $E^{(iii)}_{\rm splitting} \approx 75$~mK and $t_{\rm splitting}^{(iii)} \approx 0.1$~ns, while a larger $E_J = 5$~K yields $E^{(iii)}_{\rm splitting} \approx 6$~mK and $t_{\rm splitting}^{(iii)} \approx 1$~ns.  Still more favorable numbers appear with a modestly reduced $E_C = 0.2$~K (recall the discussion in Sec.~\ref{OptimizationStrategies}): In this case $E_J = 2$~K yields $E^{(iii)}_{\rm splitting} \approx 2$~mK and $t_{\rm splitting}^{(iii)} \approx 3$~ns while $E_J = 5$~K gives $E^{(iii)}_{\rm splitting} \approx 0.02$~mK and $t_{\rm splitting}^{(iii)} \approx 300$~ns.


We have intentionally used conservative numbers in the estimates above.  The $E_{\rm splitting}$ values that result already suggest a comfortable window between the upper and lower speed limits for the fusion-rule protocol.  We stress, however, that the splittings could be reduced by \emph{orders of magnitude}---thus widening the window further still---by, e.g., lengthening the islands and optimizing the efficacy of gate-tunable valves in the setup.
We also reiterate that the expression used in Eq.~\eqref{Eiii} likely overestimates the Coulomb contribution (see Secs.~\ref{SingleIslandSec} and \ref{OptimizationStrategies}).

\begin{figure*}
\includegraphics[width=2.05\columnwidth]{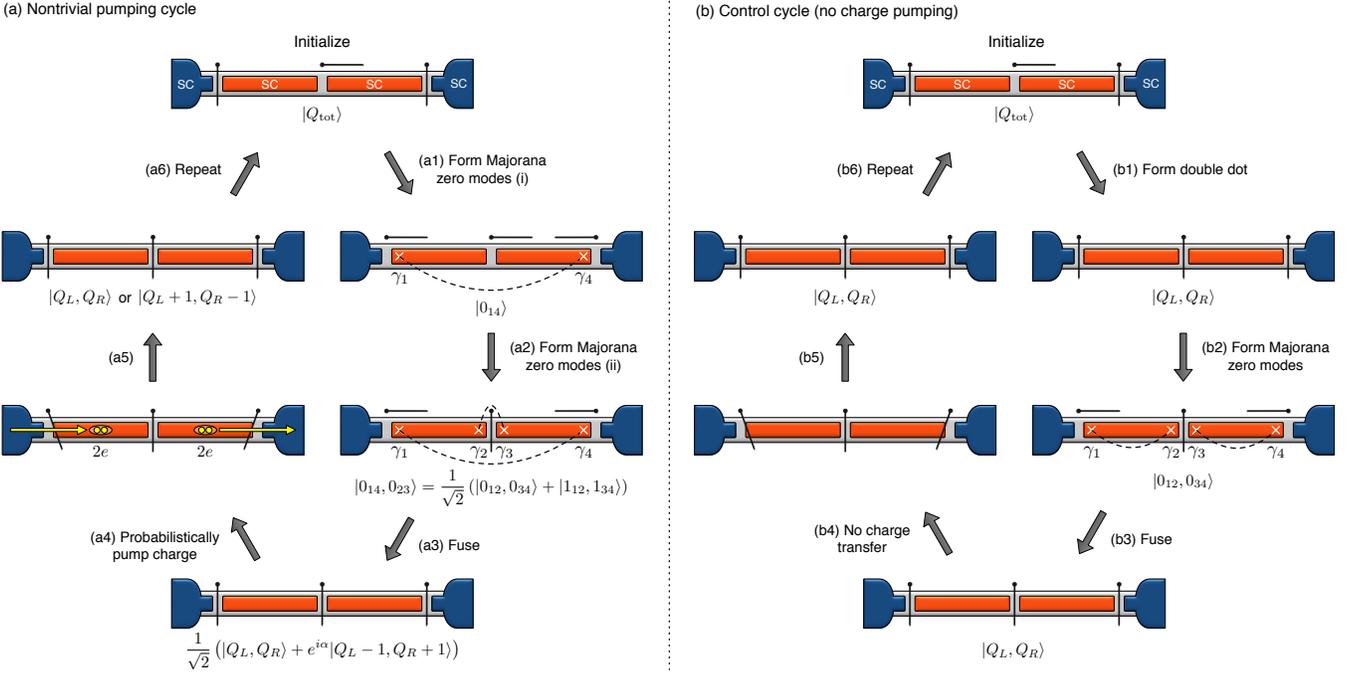}
\caption{(a) Non-trivial cycle that selectively pumps charge $2e$ whenever Ising anyons fuse into the $\psi$ channel, yielding a quantized zero-bias dc current [Eq.~\eqref{Inontrivial}] passing between the outer grounded superconductors.  The sequence initially follows the left path from Fig.~\ref{FusionRuleFig}.  Once the anyons nontrivially fuse into a superposition of $I$ and $\psi$ channels (bottom configuration) the cycle probabilistically pumps charge by pursuing the manipulations sketched in Fig.~\ref{PumpDispersions} and described in the text.  Opening the middle valve then restores the islands to their original state, whereupon the process repeats.
(b) Control cycle that initially follows the right path from Fig.~\ref{FusionRuleFig} but is otherwise identical to part (a).  This alternate sequence blocks the $\psi$ fusion channel and hence eliminates the current.  The two cycles sketched here reveal the nontrivial Ising-anyon fusion rules via a \emph{macroscopic} signal.}
\label{PumpFig}
\end{figure*}

Satisfying Eqs.~\eqref{adiabatic} through \eqref{degeneracy} is essential for implementing the charge-sensing fusion-rule experiment envisioned above.  It is also illuminating, however, to explore the limits at which the predictions break down.  Starting at rates that violate the adiabatic criterion \eqref{adiabatic} the predictions described in Sec.~\ref{IdealLimit} should be recovered as the rates are reduced, thus giving information about the system's minimal (accessible) excitation gap.  Assuming the likely scenario $t_{\rm poisoning} \gg t_{\rm splitting}$, further reducing the rate next invalidates the degeneracy criterion in Eq.~\eqref{degeneracy}.  The left and right paths in Fig.~\ref{FusionRuleFig} would then yield identical results, as in this regime the system simply follows a unique ground state throughout either protocol.  Here the crossover scale determines the precision of Majorana-related degeneracies.  This information is useful in the regime where the degeneracy is exponentially small \footnote{If such splittings are dominated by Majorana overlap within each island then we have $t_{\rm splitting} \sim e^{L/\xi}$.  Verifying this exponential scaling---e.g., by controllably modifying $\xi$---would conclusively rule out the (already unlikely) accidental Andreev bound state scenario described at the beginning of Sec.~\ref{RealisticCase}} and thus difficult to otherwise resolve \cite{AlbrechtExponential}.  Further reducing the protocol rate eventually violates Eq.~\eqref{poisoning}, thus revealing the quasiparticle poisoning time at the rate where the left and right paths again yield identical results, but with uncorrelated charge readouts of the two islands.

\subsection{Majorana-mediated charge pump: Fusion rules via transport}
\label{Pump}

Next we describe a variation on the fusion-rule-detection experiment from Sec.~\ref{FusionRuleChargeSensing} using current detection rather than charge sensing.  In short, we will cyclically fuse Ising-anyon pairs using the same sequences as before, but now selectively pump charge across the system in a fusion-channel-dependent fashion.  

\begin{figure}
\includegraphics[width=\columnwidth]{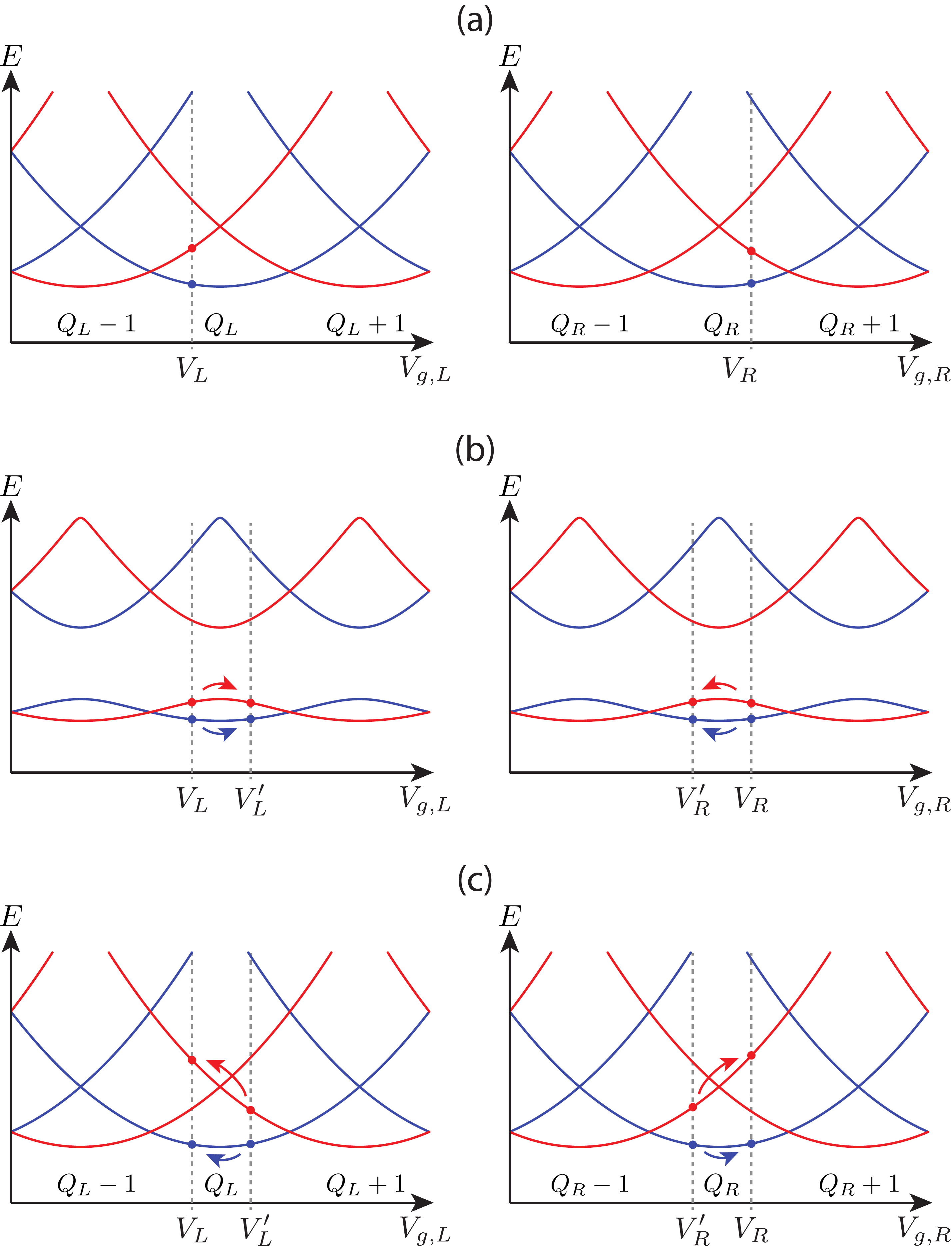}
\caption{Manipulations used to pump charge in the cycle sketched in Fig.~\ref{PumpFig}(a).  Each panel displays the energies versus gate voltage $V_{g,L/R}$ on the left/right islands.  The sequence begins from the bottom configuration of Fig.~\ref{PumpFig}(a), wherein the anyons fuse to yield a superposition of charge states $|Q_L,Q_R\rangle$ ($I$ channel) and $|Q_L-1,Q_R+1\rangle$ ($\psi$ channel).  Panel (a) denotes these charge states with circles and also indicates the initial gate voltages $V_{L/R}$.  Next one opens the outer valves, leading to the energies in panel (b), and then sweeps the gate voltages to $V_{L/R}'$.  Closing the outer valves again and subsequently restoring the initial gate voltages as shown in panel (c) completes the desired pumping: $|Q_L,Q_R\rangle$ evolves trivially while the excited state $|Q_L-1,Q_R+1\rangle$ changes to $|Q_L+1,Q_R-1\rangle$.  The latter evolution indicates charge $2e$ flowing between the outer superconductors when the fusion channel $\psi$ appears.  }
\label{PumpDispersions}
\end{figure}

Figure \ref{PumpFig}(a) depicts the nontrivial cycle of interest (see below for a companion control experiment).  The first three steps of this cycle repeat those in Fig.~\ref{FusionRuleFig}(a): Start from a charge state $|Q_{\rm tot}\rangle$, nucleate Ising anyons binding Majorana modes $\gamma_{1,\ldots,4}$, and restore charging energy to fuse them nontrivially.  Figure~\ref{PumpDispersions}(a) illustrates the energy levels of the charge states versus gate voltages $V_{g,L/R}$ for the left/right islands.  Suppose that the gate voltages are adjusted to initial values $V_{L/R}$ such that after the fusion step (a3) the system's wavefunction is $\frac{1}{\sqrt{2}}\left(|Q_L,Q_R\rangle + e^{i \alpha}|Q_L-1,Q_R+1\rangle\right)$, precisely as in \eqref{superposition}.  Again, the first and second terms in the superposition---indicated by circles in Fig.~\ref{PumpDispersions}(a)---correspond to the anyons fusing into the $I$ and $\psi$ channels, respectively.

At this stage our strategy departs from the protocol discussed in Sec.~\ref{FusionRuleChargeSensing}.
Rather than performing a charge measurement for readout, we continue with the steps shown on the left half of Fig.~\ref{PumpFig}(a) and open slightly the outer valves to Josephson couple the islands to the trivial (grounded) bulk superconductors.  The energy levels thus exhibit avoided crossings between states of equivalent charge parity as Fig.~\ref{PumpDispersions}(b) illustrates.  Next, we sequentially sweep the gate voltages to values $V_{L/R}'$, re-close the outer valves, and return the gate voltages to $V_{L/R}$ [Figs.~\ref{PumpDispersions}(b) and (c)].  After these steps the ground state $|Q_L,Q_R\rangle$ remains unchanged while the excited charge state $|Q_L-1,Q_R+1\rangle$ evolves into $|Q_L+1,Q_R-1\rangle$.  While the former outcome is trivial, the latter corresponds to the left island borrowing a Cooper pair from the left grounded superconductor and the right island analogously shedding a Cooper pair rightward.  In other words we pump charge $2e$ from left to right, \emph{but only if the anyons fuse into the $\psi$ channel}.  At this point the wavefunction collapses into either $|Q_L,Q_R\rangle$ or $|Q_L+1,Q_R-1\rangle$ with equal probability.  In either case opening the middle valve evolves the system back into the state $|Q_{\rm tot}\rangle$.  We are now back at the top of our pumping cycle in Fig.~\ref{PumpFig}(a) and ready to start anew.

Repetition of the cycle with frequency $f$ yields a time-averaged current of strength
\begin{equation}
  I_{\rm pump} = \frac{1}{2} (2e) f = ef~~~~~~(\text{nontrivial pumping cycle})
  \label{Inontrivial}
\end{equation}
flowing between the grounded superconductors.  The factor of $1/2$ accounts for the fact that the anyons fuse into a fermion $\psi$---thus contributing to the current---with only 50\% probability.  Remarkably, the fusion channel $\psi$ responsible for the presence of the excited state in Eq.~\eqref{superposition} has thus led directly to a measurable and quantized dc current at zero bias!

Applying the same sequence of operations in a different order restricts the allowed fusion outcomes and thus yields qualitatively different predictions. Consider in particular the control cycle sketched in Fig.~\ref{PumpFig}(b), which initially follows the right path in Fig.~\ref{FusionRuleFig} but is otherwise identical to Fig.~\ref{PumpFig}(a). Most importantly, step (b3) uniquely fuses the anyons back into the vacuum---deterministically yielding the ground state $|Q_L,Q_R\rangle$.  Since the $\psi$ fusion channel is unavailable, the remainder of the cycle is trivial and the time-averaged current vanishes:
\begin{equation}
  I_{\rm control} = 0~~~~~~(\text{control cycle}).
  \label{Icontrol}
\end{equation}
It is worth noting the non-Abelian nature of the experiment evident in Eqs.~\eqref{Inontrivial} and \eqref{Icontrol}.  The dependence on the order of applied external perturbations requires the topological degeneracy encoded by the Majorana zero modes and would not arise if the ground state was unique throughout the cycles (at least in the adiabatic limit).

The practical considerations outlined in Sec.~\ref{RealisticCase} apply equally well in the measurement scheme proposed here.  Essentially, the cycles in Fig.~\ref{PumpFig} should be adiabatic with respect to the minimal allowed excitation gap but swift compared to both the quasiparticle poisoning time and any inadvertent topological degeneracy splitting.  Two additional considerations are also noteworthy.  First, the gate voltage sweep $V_{L/R}' \rightarrow V_{L/R}$ in Fig.~\ref{PumpDispersions}(c) should occur faster than residual avoided crossings (not shown) between the $Q_{L/R}-1$ and $Q_{L/R}+1$ curves; this condition avoids relaxing into an unwanted charge branch.  And second, the relaxation time for resetting to a single-dot state $|Q_\mathrm{tot}\rangle$ at the top of the cycle in Fig.~\ref{PumpFig} should ideally occur on a time scale comparable to or shorter than the other steps.  A `long' relaxation time would bottleneck the cycle and limit the realizable pumped current.

Overall, operating as fast as possible within these constraints maximizes the current predicted by Eq.~\eqref{Inontrivial}. According to the time scales estimated in Sec.~\ref{RealisticCase}, a cycle frequency of $f \sim 100$~MHz easily satisfies the adiabatic criterion and yields a readily detectable current $I_{\rm pump} \sim 10$~pA.  Higher frequencies---which depending on parameters may help satisfy the lower speed limits discussed earlier---boost the signal further still.  

The cycles we introduced above share some common operating principles with the pumps studied, for example, in Refs.~\onlinecite{ChargePump,Geerligs,Niskanen,Mottonen1,Fazio,Leone,Mottonen2,Pekola}. Charging effects provide an important ingredient in all cases.  Moreover, the current orientations in such devices follow not from the sign of a bias voltage (which is not even necessary), but rather from the applied perturbations during the cycle.  We can indeed reverse the current flowing between the outer superconductors in Fig.~\ref{PumpFig}(a) simply by swapping the gate voltages applied to the left and right islands.  The central role played by the Ising-anyon fusion rules is the main feature present in our setup. We have seen that one can turn on and off the current by selectively filtering the allowed fusion channels---a large-scale manifestation of Eq.~\eqref{fusionrule}.

\section{Prototype topological qubit validation}
\label{QubitSec}

\subsection{Motivation}
\label{ValidationMotivation}

\begin{figure}
\includegraphics[width=0.9\columnwidth]{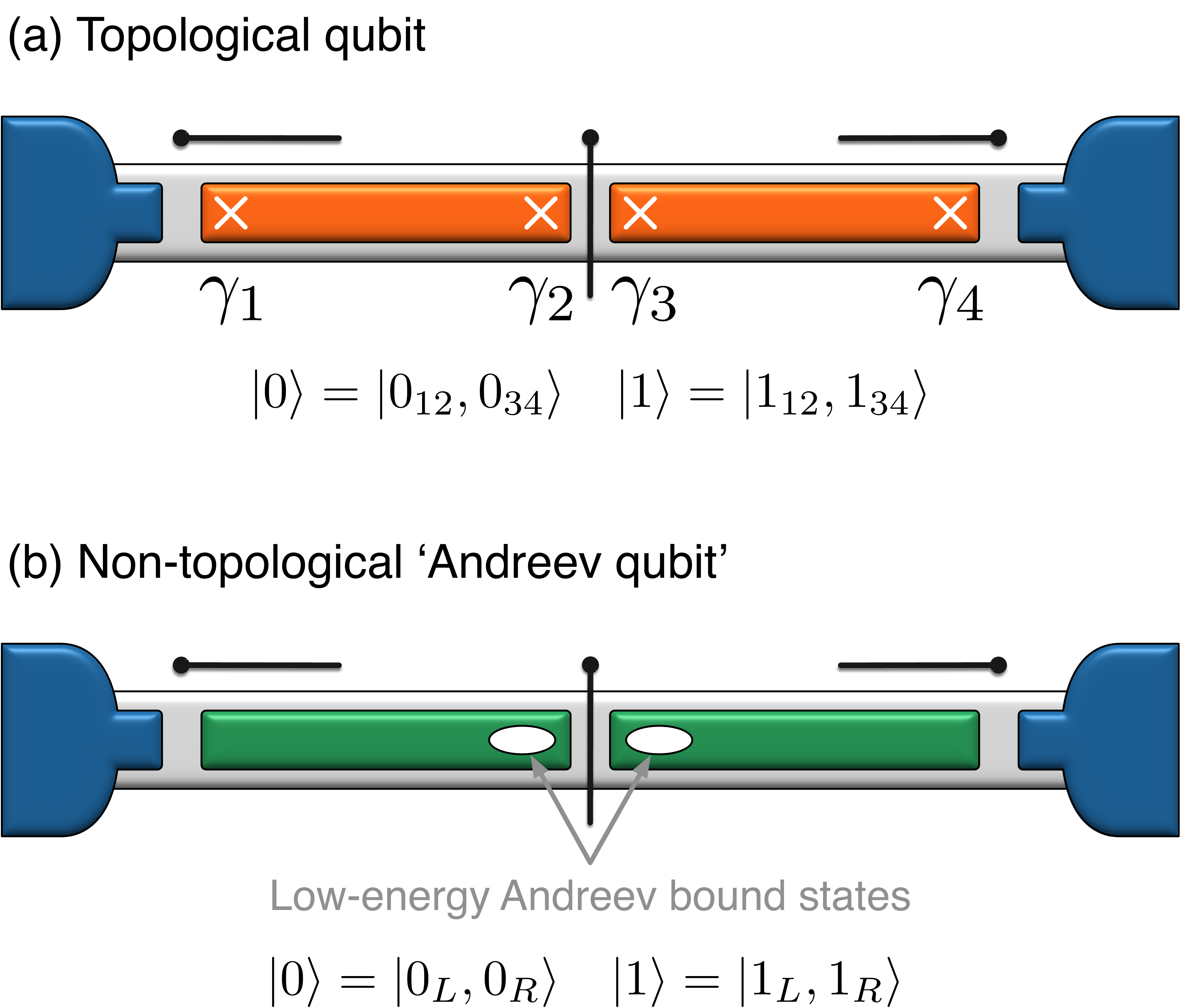}
\caption{(a) Topological qubit and (b) reference non-topological `Andreev qubit' that we seek to sharply distinguish.  For the former the qubit is stored in the topologically degenerate ground states encoded by Majorana zero modes; in the latter the occupation numbers for `accidental' near-zero-energy Andreev levels specify the qubit states.  Because of the very different sensitivity to local noise sources, we argue that the topological qubit's coherence times and oscillation frequency exhibit nontrivial scaling relations that are generically absent in the Andreev qubit and can be used to identify topological protection in a modest pre-braiding experiment.}
\label{QubitFig}
\end{figure}

The two-island setup that Sec.~\ref{FusionRuleSec} exploited for fusion-rule detection is also interesting because it constitutes a prototype topological qubit.  Assuming fixed total parity, the configuration with Majorana zero modes $\gamma_{1,\ldots,4}$ shown in Fig.~\ref{QubitFig}(a) admits two ground states that we may associate with a logical 0 and 1 \cite{NayakReview}.  It is most natural to employ the basis in Eq.~\eqref{fermions} and specifically define
\begin{equation}
  |0\rangle \equiv |0_{12},0_{34}\rangle,~~~~~|1\rangle \equiv |1_{12},1_{34}\rangle.
  \label{qubit}
\end{equation}
These states carry opposite parities within each island and can thus be read out using techniques outlined in Sec.~\ref{ReadoutSec}.

It is worth elaborating on this qubit's special properties---particularly the conditions necessary for topological protection against decoherence.  Very generally, no local measurement can read out the state of the qubit when the Majorana zero modes remain well-separated from one another; the quantum information is encoded \emph{non-locally} through the island parities.  Though certainly promising for the coherence of the qubit, it can still be prone to decoherence due to deviations from the ideal limit as shown explicitly in numerous studies \cite{GoldsteinNoise,RainisNoise,SchmidtNoise,PedrocchiNoise1,HuNoise,PedrocchiNoise2}.  For example, a stray quasiparticle entering the islands from one of the outer superconductors can take the qubit out of the computational subspace. Protection against such errors requires qubit manipulations on a time scale shorter than the typical poisoning time. Moreover, to take full advantage of the Majorana-based qubit the system should with high probability be confined to the ground-state subspace generated by the non-Abelian anyons.  This condition is satisfied if temperature, as well as the typical frequencies for environmental perturbations both fall well below the excitation gap of the system. 

In this section we assume that the above criteria are maintained and investigate the following question: How can one characterize the Majorana-based qubit in a manner that validates its topological protection?  

As part of the validation we aim to qualitatively distinguish the topological qubit in Fig.~\ref{QubitFig}(a) from the hypothetical `Andreev qubit' depicted in Fig.~\ref{QubitFig}(b).  For the latter the islands form trivial superconductors that happen to host near-zero-energy Andreev bound states (indicated by ellipses) localized near the central valve.  One could equally well encode a qubit in such a system using the occupation numbers $n_{L/R}$ of the left/right Andreev modes, i.e., $|0\rangle \equiv |0_L,0_R\rangle$ and $|1\rangle \equiv |1_L,1_R\rangle$.  The two qubits in Fig.~\ref{QubitFig} in fact share several common features despite their different physical origin.  In either case, one can introduce (not topologically protected) qubit rotations by opening the middle valve for a specified time; readout may be performed using charge-sensing protocols; and both qubits appear similarly susceptible to errors from poisoning events.  Exposing the unique characteristics that stem from the non-local nature of the topological qubit therefore requires some care \footnote{In principle one could use the schemes proposed in Refs.~\onlinecite{Burnell1,Burnell2} to distinguish the topological and Andreev qubits by employing \emph{local} charge-sensing techniques \cite{Gilad} that probe the rather different charge distributions in the two setups.  Our proposal outlined below can be viewed as extending these interesting studies by illuminating the topological characteristics in an experiment that invokes less ambitious \emph{global} charge sensing on the islands.}. 

The fusion-rule experiments described previously already shed some light on the issue.  Indeed the different paths in Fig.~\ref{FusionRuleFig} rigidly initialize the topological qubit into either $|0\rangle$ (right path) or $(|0\rangle +|1\rangle)/\sqrt{2}$ (left path); reproducing the same initializations for the Andreev qubit, by contrast, would require fine-tuning as discussed at the end of Sec.~\ref{IdealLimit}.  As a much more compelling demonstration one could exchange a pair of Majorana zero modes to implement the fault-tolerant qubit rotations generated by the non-Abelian braid matrices.  Certainly no analog exists for the trivial Andreev qubit.  While we do explore braiding in Sec.~\ref{BraidingSec}, such an experiment requires departing from the single-wire geometry of Fig.~\ref{QubitFig}---which we wish to presently avoid.  

Instead we will argue that one can meaningfully contrast the qubits in Fig.~\ref{QubitFig} through the behavior of their coherence times $T_1$ and $T_2$, which can be extracted via non-topological manipulations.  To obtain precise experimental predictions we focus on how these quantities scale with the (experimentally tunable) qubit oscillation frequency $\omega_0$ stemming from residual average splittings of the $|0\rangle$ and $|1\rangle$ energies.  Assuming decoherence is primarily caused by low-frequency environmental noise in the low-energy subspace, we obtain characteristic scaling relations that link $1/T_1$, $1/T_2$, \emph{and} $\omega_0$ for the topological setup.

Before turning to details we sketch an intuitive picture for the origin of these scaling relations.  Suppose for concreteness that the oscillation frequency $\omega_0$ arises predominantly from overlap of Majorana zero modes across a given superconducting island, which scales as $e^{-L/\xi}$ (up to an important oscillatory prefactor discussed below; as usual $L$ denotes the island size while $\xi$ is the induced correlation length).  Environmental fluctuations stochastically modify the Majorana overlap and thus the degeneracy splitting---thereby dephasing the topological qubit.  Crucially, since \emph{all} local noise sources affect the instantaneous splitting through $e^{-L/\xi}$ (or the oscillatory prefactor), such noise-induced energy variations are bounded by the same exponential envelope as the oscillation frequency $\omega_0$ itself \footnote{If noise changes $\xi \rightarrow \xi+ \delta \xi(t)$, with small-amplitude fluctuations $\delta \xi(t)$, then the degeneracy splitting becomes time-dependent and scales as $e^{-L/[\xi + \delta \xi(t)]} = e^{-L/\xi}\left(1+ \frac{L \delta \xi}{\xi^2} + \cdots\right)$.  Other parameters of course vary as well, which can change the oscillatory prefactor leading to additional interesting physics elaborated below.}.  Hence, \emph{fluctuations and time-averaged quantities are linked for the topological qubit}.  This property relates $\omega_0$ to the dephasing time $T_2$ and applies quite generally provided the topological degeneracy holds within exponential accuracy.  The relaxation time $T_1$ of the topological qubit also varies with the average splitting $\omega_0$, thus linking $T_1$ and $T_2$ as well.  We stress that such relations do not hold for the non-topological Andreev qubit; there purely local noise sources induce stochastic energy variations that need not bear any relation to the time-averaged bound state energies.  

\subsection{Model for noise in a topologically protected qubit}

As discussed above, we assume that the relevant subspace consists solely of logical qubit states $|0\rangle = |0_{12},0_{34}\rangle$ and $|1\rangle = |1_{12},1_{34}\rangle$.  We therefore model noise with the following time-dependent Hamiltonian:
\begin{equation}
  H(t) = h_z(t) \sigma^z + h_x(t) \sigma^x.
  \label{Hqubit}
\end{equation}
In the basis specified above the Pauli matrix $\sigma^z$ is diagonal whereas $\sigma^x$ flips the qubit.

We consider two physical contributions to the qubit coupling $h_z(t)$: $(i)$ Overlap of the Majorana zero modes across a given island in Fig.~\ref{QubitFig}(a) produces a degeneracy splitting $\propto \cos(\kappa L)e^{-L/\xi}$, where the parameter $\kappa$ is related to the wire's effective Fermi wavevector~\cite{SmokingGun}. Both $\kappa$ and the correlation length $\xi$ depend on microscopic parameters such as the Zeeman splitting and the nanowire's chemical potential. $(ii)$ Residual charging energy of the islands yields a splitting exponentially small in $E_J/E_C$, with a prefactor $\propto (E_C E_J^3)^{1/4}\cos(\pi n_0)$ [see Eq.~\eqref{ECsplitting}]. These two mechanisms weakly lift the topological degeneracy and thus energetically split the $|0\rangle$ and $|1\rangle$ qubit states.

The $h_x(t)$ coupling arises from hybridization of Majorana zero modes $\gamma_{2}$ and $\gamma_{3}$ residing opposite the central valve and is exponentially small in $W/\xi_U$, where $W$ is the width separating the islands and $\xi_U$ is the Majorana decay length into the barrier region; see the discussion near Eq.~\eqref{Eii}.  Beginning from a $\sigma^z$ eigenstate, this coupling tends to rotate the qubit between the $|0\rangle$ and $|1\rangle$ states.

Fluctuations in nearby gates, the applied magnetic field, etc.\ stochastically impact the parameters $\kappa$, $\xi$, $E_J$, $E_C$, $n_0$, and $\xi_U$, thus generating time dependence in the qubit couplings.  The specific noise source is not especially important for our analysis, since for the topological qubit there exist no local noise sources which lead to an instantaneous splitting not involving an exponential.  The minimal model defined here allows us to quantify how the intimate connection between time-averages of these couplings and their typical fluctuations affect the topological qubit's properties.

\begin{figure}
\includegraphics[width=0.8\columnwidth]{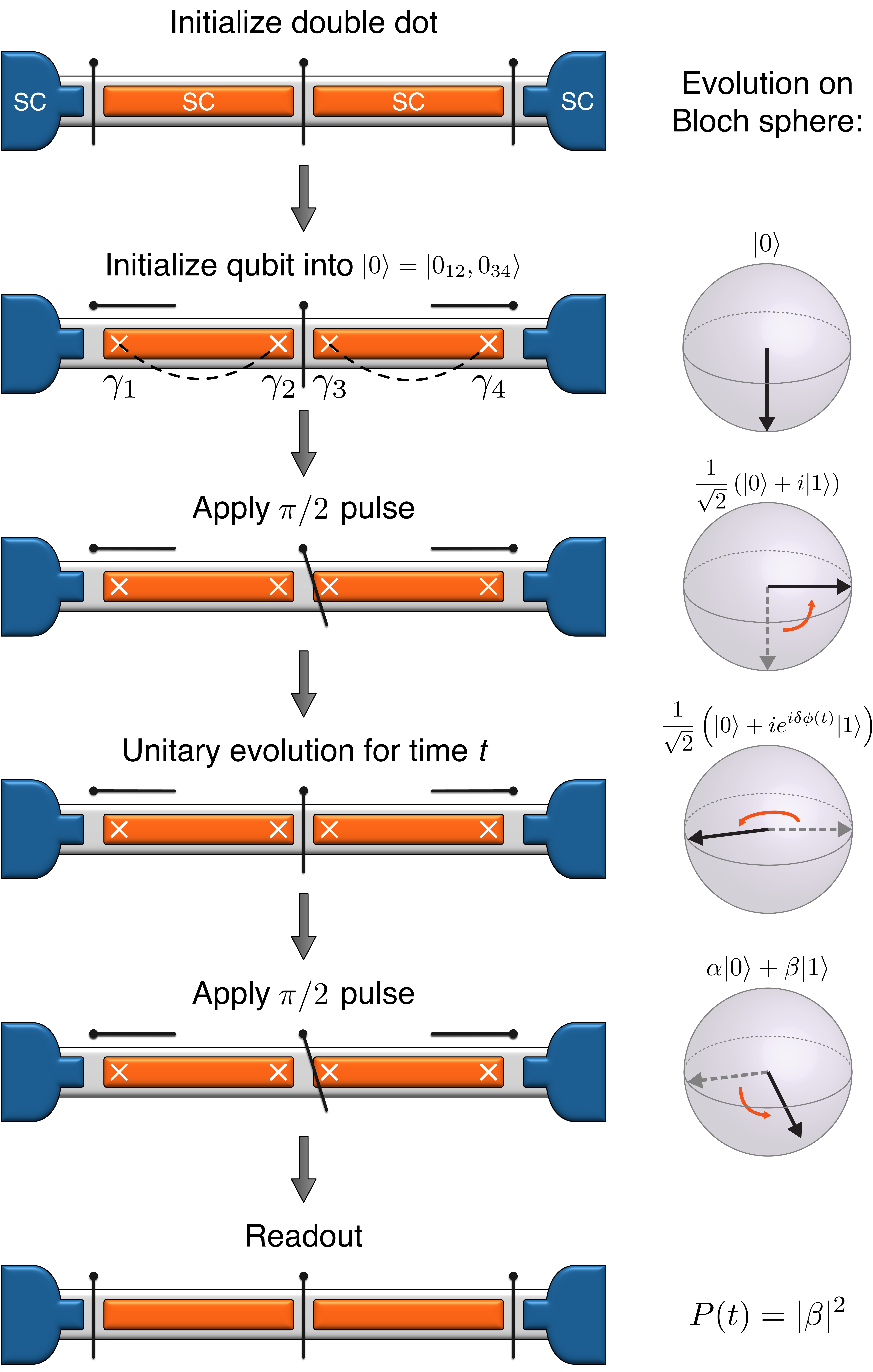}
\caption{Protocol for measuring the topological qubit's dephasing time $T_2$ and oscillation frequency $\omega_0$.  The left side illustrates the valve manipulations while the right shows the corresponding qubit orientation on the Bloch sphere.  After initialization and a $\pi/2$ pulse, the qubit unitarily evolves for time $t$.  During this evolution, the relative phase between $|0\rangle$ and $|1\rangle$ changes by an amount $\delta \phi(t)$ set by $\omega_0$ and time-dependent noise in the  splitting of the qubit states.
Crucially, the latter two factors are intimately related for the topological qubit because they are constrained by the same exponential suppression.
After a second $\pi/2$ pulse, the probability of measuring the $|1\rangle$ state is $P(t) = |\beta|^2 = \frac{1}{2}\left[1 + \cos\delta\phi(t)\right]$.  Noise-averaging gives Eq.~\eqref{Pt2} which simultaneously probes both $T_2$ and $\omega_0$.  The nontrivial connection between uniform quantities and fluctuations for the topological qubit implies the unusual scaling relation $\omega_0 \sim 1/T_2$ that signifies topological protection.  A similar scaling relation links $T_1$ and $T_2$; see Eq.~\eqref{scaling2}.}
\label{T2Fig}
\end{figure}

\subsection{Dephasing time $T_2$}
\label{T2sec}

We focus first on dephasing of the topological qubit.  For simplicity we suppose that the $\sigma^z$ coupling dominates, which seems reasonable given the closed middle valve in Fig.~\ref{QubitFig}(a).  We will thus set $h_x(t) = 0$ unless stated otherwise and parametrize the remaining coupling via
\begin{equation}
  h_z(t) = \frac{\hbar \omega_0}{2} + \delta h_z(t).
\end{equation}
Here $\omega_0$ denotes the qubit oscillation frequency while $\delta h_z(t)$ encodes noise.

One can probe dephasing using the following standard manipulations sketched in Fig.~\ref{T2Fig}: (1) Initialize the qubit into the state $|0\rangle$.  (2) Apply a $\pi/2$ pulse $U_{\pi/2} = e^{i (\pi/4) \sigma^x}$ by opening the middle valve to turn on the $h_x$ coupling in Eq.~\eqref{Hqubit} for a prescribed duration, thus orienting the qubit on the equator of the Bloch sphere.  (Note that the Majorana wavefunctions do not oscillate in the barrier between the islands, which facilitates the controlled rotation of the qubit \cite{SauWireNetwork}.)  (3) Let the system evolve for a time $t$.  (4) Apply a second $\pi/2$ pulse.  After this sequence the probability of measuring the qubit in state $|1\rangle$ reads
\begin{equation}
  P(t) = \frac{1}{2} \left[1+\cos\left(\omega_0 t + \frac{2}{\hbar}\int_0^t dt' \delta h_z(t')\right)\right].
  \label{Pt}
\end{equation}

Following closely Ref.~\onlinecite{Koch}, we now investigate how $P(t)$ behaves upon averaging over noise, i.e., when averaging over many measurements.  We take $\delta h_z$ to obey Gaussian correlations with $\langle \delta h_z(t)\rangle = 0$ and $\langle \delta h_z(t) \delta h_z(t')\rangle = S_z(t-t')$, where $S_z(t-t')$ determines the noise power spectrum. We will assume low-frequency $1/f$ noise to be dominant so that
\begin{equation}
  S_z(\omega) \sim \frac{(\Delta E_{\rm typ}^z)^2}{|\omega|},
  \label{NoisePower}
\end{equation}
where $\Delta E_{\rm typ}^z$ sets the typical fluctuation-induced variation in the $\sigma^z$ coupling.  Equation~\eqref{NoisePower} applies for frequencies $\omega$ between an `infrared' cutoff $\Lambda_{\rm min}$ and an `ultraviolet' cutoff $\Lambda_{\rm max}$ given by the bulk gap.  Noise-averaging the probability in Eq.~\eqref{Pt} then yields \cite{Koch}
\begin{equation}
  \langle P(t)\rangle \approx \frac{1}{2} \left[1+\cos(\omega_0 t)e^{-(\Delta E_{\rm typ}^z t/\hbar)^2 f(t)}\right].
  \label{Pt2}
\end{equation}
In the limit $\Lambda_{\rm min} t \ll 1$ and $\Lambda_{\rm max} t \gg 1$ the function $f(t)$ above depends on time only logarithmically.  Ignoring this weak dependence, we thus obtain a dephasing time
\begin{equation}
  T_2 \sim \frac{\hbar}{\Delta E_{\rm typ}^z}.
  \label{T2}
\end{equation}

Similar behavior is of course valid for any type of qubit.  Again, the special property of the topological qubit is that $\omega_0$ and the typical fluctuations $\Delta E_{\rm typ}^z$ are linked through the exponential suppression of the ground-state splitting, which leads to our first scaling relation,
\begin{equation}
  \omega_0 \sim \frac{1}{T_2} \sim \left\{
                \begin{array}{ll}
                  e^{-L/\xi},& {\rm Majorana~overlap~dominant} \\
                  e^{-\sqrt{8E_J/E_C}},& {\rm charging~energy~dominant}.
                \end{array}
              \right.
   \label{scaling}
\end{equation}
Thus as the topological qubit becomes `perfect', both $\omega_0$ and $1/T_2$ vanish in a correlated fashion.  The right side indicates more precisely how these quantities scale in the limits where Majorana overlap and weak residual charging effects dominate the topological degeneracy splitting; note that the prefactors suppressed in Eq.~\eqref{scaling} exhibit sub-exponential dependence on parameters.  In practice one can explore this relation by varying $\xi$ with, e.g., a back gate, or by modifying $E_J/E_C$ through modulation of the outer valves in Fig.~\ref{QubitFig}(a).  

In the Majorana-overlap-dominated regime, the scaling relation in Eq.~\eqref{scaling} emphasizes the exponential envelope of the degeneracy splitting that is central to topological protection.  As noted above, however, here the time-averaged splitting $\omega_0$ \emph{also} includes an oscillatory prefactor $\cos(\kappa L)$ with quite interesting physical consequences.  For instance, when $\omega_0$ becomes extremal, $\delta h_z$ is much less sensitive to noise in $\kappa$ if the typical fluctuations in $\kappa L$ are small---thus suppressing $1/T_2$.  When $\omega_0$ is instead tuned near zero, $\delta h_z$ is maximally sensitive to noise in $\kappa$ leading to enhancement of $1/T_2$.  One thereby obtains out-of-phase oscillations in $\omega_0$ and $1/T_2$ when changing the system's parameters---another key consequence of the unconventional link between time-averages and fluctuations for the qubit.  Exploring refinements of the scaling behavior on top of the dominant exponential $e^{-L/\xi}$ thus reveals additional fingerprints of Majorana modes that provide further evidence of topological protection.  We stress that the oscillations described above may actually be easier to observe than the exponential itself since they should be visible even with modest parameter changes that only weakly impact $L/\xi$.

A few remarks are in order to put our discussion in proper perspective: 

First, the precise scaling between $\omega_0$ and $T_2$ depends on the noise model used.  For example, with short-range-correlated noise it is straightforward to show, again following Ref.~\onlinecite{Koch}, that $\omega_0 \sim 1/\sqrt{T_2}$ and hence both quantities vanish exponentially as the qubit becomes perfect.  The important universal feature is that the dephasing rate exhibits some characteristic dependence on the typical fluctuation amplitude for the $\sigma^z$ coupling in Eq.~\eqref{Hqubit}; that alone is sufficient to link $\omega_0$ and $T_2$ via a scaling relation akin to Eq.~\eqref{scaling}.

Second, the entirety of Eq.~\eqref{scaling} (or a straightforward extension if the noise is not $1/f$-like) needs to be satisfied to verify topological protection according to the ideas in this section.  That is, $\omega_0$ and $1/T_2$ must scale with one another, \emph{and} they must both scale exponentially with some parameter.  Merely demonstrating $\omega_0\propto1/T_2$ is not sufficient.  This weaker relation in fact arises in the seminal spin qubit work of Petta et al.~\cite{Petta05}, where topological physics is not operative; in that case, exponential scaling is absent.

Finally, for realizing our predicted scaling relations, it is most certainly not sufficient to simply have a qubit with an exponentially small time-averaged splitting in some parameter. 
To illustrate this important point, consider two localized spin-1/2 particles whose orbital wavefunctions decay exponentially on a length scale $\xi_s$ [cf.~the example discussed after Eq.~\eqref{fusionrule}].  When these spins are infinitely far apart and with all time-averaged external fields etc.~`turned off', the singlet and triplet states for the two-spin system are exactly degenerate.  Moreover, at large but finite spatial separation $L_s$, these states split by a time-averaged amount $\omega_0\propto e^{-L_s/\xi_s}$, i.e., this spin qubit exhibits an exponential $\omega_0$ scaling quite similar to our topological qubit.  For the spin qubit, however, local noise sources that time-average to zero produce \emph{additional terms} in the instantaneous splitting that are by no means constrained by the above exponential---the most obvious example being stray Zeeman fields acting separately on each spin.  These additional local noise sources yield a dephasing rate $1/T_2$ that violates the exponential scaling relation in Eq.~\eqref{scaling}.  Conversely, \emph{all} local noise sources present for our topological qubit, including fluctuations in the Zeeman field, feed into the prefactor and/or exponent of the exponentially dependent instantaneous splitting, e.g., $\cos(\kappa L)e^{-L/\xi}$.  Hence, the dephasing rate due to all local noise sources can be exponentially suppressed by changing a single dimensionless parameter, and Eq.~\eqref{scaling} is fully satisfied.  The part of Eq.~\eqref{scaling} involving $T_2$ thus verifies, in a rather general and universal fashion, the topological qubit's inherent non-locality.

\subsection{Relaxation time $T_1$}
\label{T1}

We now briefly address the time scale $T_1$ at which the topological qubit relaxes towards its thermodynamic equilibrium state; a more in-depth discussion can be found in Appendix~\ref{T1appendix}.  Suppose for simplicity that we work in the temperature regime $\hbar\omega_0 \ll k_B T \ll {\rm bulk~gap}$ (which is not unreasonable) and that relaxation is induced by classical noise in weak $\sigma^x$ coupling $h_x(t)\ll \hbar \omega_0$ in Eq.~\eqref{Hqubit}.  If one initializes the system into, say,  the $|0\rangle$ logical qubit state (which is split from the $|1\rangle$ state by a finite energy $\hbar\omega_0$) via the manipulations in the top two panels of Fig.~\ref{T2Fig}, the $h_x(t)$ coupling will subsequently induce `bit flips' and hence cause the initial state to decay in the presence of noise.  Given the assumptions specified above, measuring the qubit at later times $t \gg T_1$ will yield $|0\rangle$ or $|1\rangle$ with equal probability.

Assuming low-frequency-dominated noise, the relaxation time $T_1$ generally \emph{decreases} upon reducing $\omega_0$, i.e., the system relaxes more quickly as the topological degeneracy splitting tends to zero.
 This trend arises because smaller $\omega_0$ renders random $\sigma^x$ bit flip processes more efficient. One can thus probe $\omega_0$ not only from the oscillations in Eq.~\eqref{Pt2} but also by probing $T_1$.  Appendix~\ref{T1appendix} evaluates $T_1$ in a simplified model (see also Ref.~\onlinecite{Slichter}) with $h_x(t)$ exhibiting $1/f$ noise of typical amplitude $\Delta E_{\rm typ}^x$ and specifically finds
\begin{equation}
  T_1 \sim \frac{\hbar^2\omega_0}{(\Delta E_{\rm typ}^x)^2}.
  \label{T1main}
\end{equation}
Note that since we assume $\omega_0 \gg \Delta E_{\rm typ}^x$, the $T_1$ time is still exponentially long despite the $\omega_0$ dependence in the numerator. Combining Eqs.~\eqref{scaling} and \eqref{T1main} leads to our second scaling relation, 
\begin{equation}
  T_1(\omega_0) \propto \frac{1}{T_2(\omega_0)},
  \label{scaling2}
\end{equation}
which relates the dependence of the two coherence times on the ground-state splitting encoded in $\omega_0$ and holds at fixed $\Delta E_{\rm typ}^x$.
In practice it should be possible to vary $\omega_0$ nearly independently from $\Delta E_{\rm typ}^x$ and hence probe this scaling relation; for instance, modulating the outer valves in Fig.~\ref{QubitFig}(a) could tune the former while impacting the latter very little.  We stress that Eq.~\eqref{scaling2} relates information about the \emph{uniform} oscillation frequency $\omega_0$ encoded in $T_1$ to information about \emph{fluctuations} encoded in $T_2$.  This behavior together with Eq.~\eqref{scaling} provide fingerprints of topological protection observable in a relatively modest single-wire experiment.

\section{Braiding and non-Abelian statistics}
\label{BraidingSec}

\subsection{Preliminaries}

Demonstration of fusion-rule detection and topological qubit validation as outlined earlier would provide strong evidence for the protection of quantum information \emph{stored} in Majorana-based qubits.  This section introduces an all-electrical braiding protocol that allows topologically protected qubit \emph{manipulations}, thereby stepping further towards applications. We continue to focus on minimal geometries sufficient to establish the desired effect in a controlled experiment difficult to emulate with non-topological means.  Specifically, we seek to demonstrate `rigid' rotation of a single topological qubit under an exchange, which lies at the heart of non-Abelian statistics and topological quantum information processing.

\begin{figure*}
  \includegraphics[width=\textwidth]{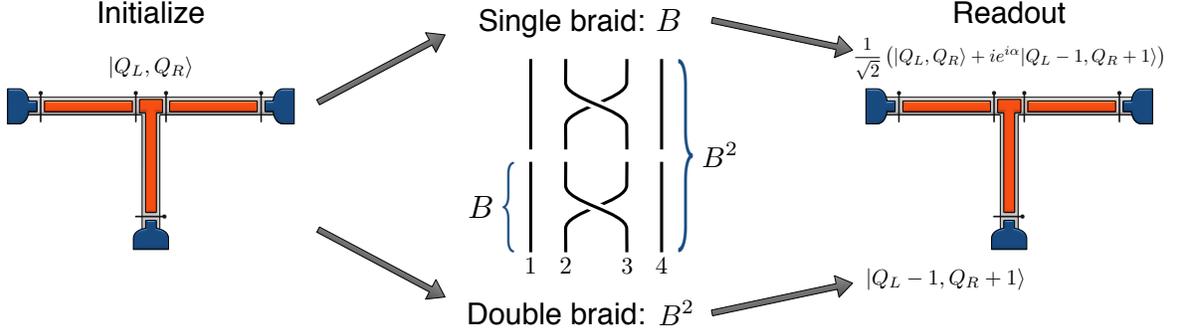}
  \caption{(a) Elementary braid of Majorana zero modes in a trijunction hosting a single topological qubit.  Modulating the three valves adjacent to the vertical island controls both charging effects and the coupling between neighboring topological superconductors.  Steps (1) through (3) use this flexibility to braid the inner Majorana modes $\gamma_2$ and $\gamma_3$, thus nontrivially rotating the topological qubit via the non-Abelian braid matrix in Eq.~\eqref{BraidMatrix}.  (b) Experimental protocols for verifying `rigid' qubit rotation induced by braiding.  Beginning from the ground state of decoupled, charging-energy-dominated islands, the topological qubit is initialized into the $|0\rangle$ state by opening the outer valves.  A single braid (upper path) or double braid (lower path) rotates the topological qubit, and then charging energy is restored for readout.  Under the single braid the qubit rotates into an equal superposition of $|0\rangle$ and $|1\rangle$ leading to maximally uncertain measurement outcome; the double braid flips the qubit to $|1\rangle$ and returns a unique measurement value.    }
  \label{BraidingFig}
\end{figure*}

For this purpose we must abandon the elegant single-wire geometries and turn to networks \cite{AliceaBraiding,ClarkeBraiding,SauBraiding,HalperinBraiding,BraidingWithoutTransport,BeenakkerBraiding,BondersonBraiding}.  Throughout this section we consider the T-junction sketched in Fig.~\ref{BraidingFig}.  Three superconducting islands coat the junction and are separated from each other and outer bulk superconductors by gate-tunable valves.  One can view the two islands on the horizontal leg as furnishing the topological qubit, very similar to Fig.~\ref{QubitFig}(a), that we will manipulate by braiding Majorana zero modes with the aid of the vertical island.  [Note that if the inner valves in the figure sit too close to one another, operating them independently may prove technically challenging.  The protrusion of the vertical island onto the horizontal leg alleviates this issue due to screening from the intervening superconducting segment.  Its width should be smaller than the correlation length to avoid trivially decoupling the left and right islands (a length scale of $\sim200$~nm or so seems reasonable).]

In what follows we first introduce our scheme for modulating valves to effect an elementary braiding operation, and then turn to more detailed protocols for initialization, rotation, and readout of the topological qubit.  All manipulations discussed hereafter should be carried out slowly relative to the minimal excitation gap for the system but fast compared to residual degeneracy splittings and quasiparticle poisoning times; for details see Ref.~\onlinecite{Hell} as well as the related discussion in Sec.~\ref{RealisticCase}.

\subsection{Elementary braid}
\label{ElementaryBraidSec}

The sequence of operations comprising the single braid of interest appears in Fig.~\ref{BraidingFig}(a).  In the leftmost configuration the two decoupled horizontal islands are both topological and together host four Majorana modes $\gamma_{1,\ldots,4}$.  (The essentially isolated vertical island contributes no zero modes at this stage since charging energy dominates there.)  At fixed parity the system therefore supports two degenerate ground states that encode one logical qubit.  In terms of occupation numbers for fermions $f_{12} = (\gamma_1 + i \gamma_2)/2$ and $f_{34} = (\gamma_3+i\gamma_4)/2$ defined in Eq.~\eqref{fermions}, we take these states to be
\begin{equation}
  |0\rangle \equiv |0_{12},0_{34}\rangle,~~~~~|1\rangle \equiv |1_{12},1_{34}\rangle,
\end{equation}
precisely as in Sec.~\ref{QubitSec}.  Our goal is to electrically braid the inner Majorana modes $\gamma_{2,3}$ by operating the valves in Fig.~\ref{BraidingFig}(a), and to recall how the qubit transforms as a result.

One can distill the braid into three steps summarized below and in Fig.~\ref{BraidingFig}(a).  For brevity let $L$, $R$, and $V$ respectively denote the left, right, and vertical islands in the trijunction.

(1) Open both the left and bottom valves adjacent to island $V$. Opening these valves extends the topological phase for island $L$ down into the vertical leg---whose charging energy is then quenched by strong Josephson coupling to the adjacent bulk superconductor as well as coupling to island $L$.  Majorana mode $\gamma_2$ thereby shuttles to the bottom end of island $V$.  (Precisely how one opens the valves is unimportant.  One should, however, avoid completely opening the bottom valve while the left valve remains closed since $V$ would then host an additional, unwanted pair of Majorana zero modes. This spurious degeneracy would spoil the topological nature of the manipulation.  Similar remarks hold for the subsequent steps outlined below.)

(2) Close the left valve adjacent to island $V$ while opening the right one, thus transporting $\gamma_3$ from island $R$ to $L$.  The topological superconductors have now been `resewn' such that $V$ and $R$ form a continuous topological phase.

(3) Close all valves adjacent to $V$ to restore dominant charging energy for that island.   This operation transfers $\gamma_2$ to island $R$ [analogously to step (1)] and completes the braid. 

We note that the bottom bulk superconductor in Fig.~\ref{BraidingFig} is strictly speaking unnecessary for this braiding protocol, though it is practically useful.  Without this component the charging energy for the lower island can still be quenched if inter-island Josephson coupling overwhelms its charging energy.  Weaker inter-island Josephson couplings can, however, be tolerated if charging energy is quenched by strong Josephson coupling to the bottom bulk superconductor, thus relaxing the operating requirements for some of the valves.

The above sequence exchanges `half' of the fermion $f_{12}$ with `half' of $f_{34}$ and thus quite naturally impacts the qubit in a dramatic way.  More precisely, Refs.~\onlinecite{AliceaBraiding,ClarkeBraiding,SauBraiding,HalperinBraiding,BondersonBraiding} show from various viewpoints that the qubit undergoes unitary rotation specified by the braid matrix
\begin{equation}
  U_{\rm braid} = e^{\pm i \frac{\pi}{4}\sigma^x}.
  \label{BraidMatrix}
\end{equation}
The operator $\sigma^x$ in Eq.~(\ref{BraidMatrix}) flips the qubit state while the sign in the exponent depends on details of the trijunction \cite{AliceaBraiding,ClarkeBraiding,SauBraiding}.
(This sign does not affect the measurement probabilities in the protocols discussed below, but we retain it for completeness.)
Interestingly, essentially the same braiding properties arise for quasiparticles in the non-Abelian Moore-Read quantum Hall state and vortices in 2D topological superconductors \cite{MooreRead,ReadGreen,Ivanov}.

One may recall that precisely the same $\pi/2$ qubit rotation encoded in Eq.~\eqref{BraidMatrix} was invoked in Sec.~\ref{QubitSec} for topological qubit validation, but via \emph{unprotected} manipulations that required fine-tuning.  We stress that Eq.~\eqref{BraidMatrix} on the contrary represents the rigid rotation that we are interested in demonstrating; it requires no fine-tuning and is exact in the limit where the topological degeneracy is perfect and the system remains in the ground-state manifold throughout the braid.

\subsection{Experimental protocols}

We now propose a series of experiments designed to probe the braid matrix in Eq.~\eqref{BraidMatrix}.  Consider first the initialize $\rightarrow$ single braid $\rightarrow$ readout protocol sketched in the upper part of Fig.~\ref{BraidingFig}(b).  The starting point is a `triple dot' configuration where all valves in the trijunction are closed, leaving three isolated islands dominated by charging energy.  We let the system relax into the ground state $|Q_L,Q_R\rangle$ with charges $Q_{L/R}$ on the left/right islands (charge on the vertical island is unimportant and thus suppressed).  Subsequently opening the outermost left and right valves nucleates our four Majorana modes and initializes the topological qubit into the logical $|0\rangle$ state.  This stage of the protocol thus sends
\begin{equation}
  |Q_L,Q_R\rangle \rightarrow |0\rangle~~~~~({\rm initialization}).
\end{equation}
Next, the inner Majorana modes $\gamma_{2,3}$ undergo a single braid as outlined in Sec.~\ref{ElementaryBraidSec}, transforming the qubit according to
\begin{equation}
  |0\rangle \rightarrow U_{\rm braid} |0\rangle = \frac{|0\rangle \pm i |1\rangle}{\sqrt{2}}~~~~~({\rm single~braid}).
\end{equation}
Finally, for readout purposes we re-close the outer valves to convert the degenerate qubit states into non-degenerate charge eigenstates.  State $|0\rangle$ evolves back to $|Q_L,Q_R\rangle$ while we assume that qubit state $|1\rangle$ evolves into $|Q_L-1,Q_R+1\rangle$, i.e.,
\begin{equation}
  \frac{|0\rangle \pm i |1\rangle}{\sqrt{2}} \rightarrow \frac{|Q_L,Q_R\rangle \pm i e^{i\alpha}|Q_L-1,Q_R+1\rangle}{\sqrt{2}}~~~~~({\rm readout})
  \label{readout}
\end{equation}
for some non-universal $\alpha$.  Measurement through either charge readout or cyclic charge pumping (recall Secs.~\ref{ReadoutSec} and \ref{FusionRuleSec}) should then detect $|0\rangle$ or $|1\rangle$ with equal probability.  This probabilistic result indicates that the qubit transforms nontrivially, despite the fact that the protocol returns the device to its precise starting configuration!

We must nevertheless exercise caution: Exactly the same outcome would arise if, say, charges hopped randomly between the two horizontal islands during the protocol, even if one failed to braid.  Fortunately control experiments can rule out uninteresting noise-related origins of such behavior.  For example, suppose that instead of braiding we simply transport Majorana modes from the horizontal islands to the vertical island and then back again [e.g., complete step (1) from Fig.~\ref{BraidingFig}(a) and then immediately undo it].  Although the Majorana modes shuttle along the trijunction in a similar fashion, the absence of a braid should return the topological qubit deterministically to its original state.  A successful warm-up experiment of this type would demonstrate controllable transfer of Majorana modes between islands in the trijunction.  Moreover, observing the predicted distinction between the braid and such control protocols would support the qubit transformation having its roots in non-Abelian statistics.

The initialize $\rightarrow$ double braid $\rightarrow$ readout protocol from the lower end of Fig.~\ref{BraidingFig}(b) provides an even more compelling control.  In a straightforward modification of the single-braid experiment, here we obtain
\begin{align}
{} & {} |Q_L,Q_R\rangle \rightarrow |0\rangle & & ({\rm initialization})
  \nonumber \\
 {} & {} |0\rangle \rightarrow (U_{\rm braid})^2 |0\rangle = \pm i |1\rangle & & ({\rm double~braid})
  \\
 {} & {} |1\rangle \rightarrow |Q_L-1,Q_R+1\rangle & & ({\rm readout}).
   \nonumber
\end{align}
Whereas singly braiding yields a maximally uncertain measurement outcome, a double braid flips the qubit with 100\% probability so that measurement becomes deterministic.  Confirming this remarkable behavior would not only definitively rule out random charge fluctuations as the source of qubit rotation under a single braid, but also verify the `rigidity' of the braid matrix in Eq.~\eqref{BraidMatrix}.

Regarding the latter point, we note that our measurement scheme is not sensitive to the relative phase of $\pm i$ between $|0\rangle$ and $|1\rangle$ in Eq.~\eqref{readout}.  Equal measurement outcomes merely imply that a braid rotates the qubit by $\pi/2$ about \emph{some} axis normal to the poles of the Bloch sphere.  In general, the double-braid control experiment deterministically flips the qubit only if each individual braid implements a $\pi/2$ rotation about the \emph{same} axis.  Deviations from perfect rigidity can, however, arise from imperfections such as dephasing during the protocol and residual splitting of the ground-state degeneracy.  Investigating the precision of braid transformations experimentally should allow one to further validate the topological qubit beyond the means described earlier.

\section{Outlook}
\label{OutlookSec}

Successful completion of each milestone explored in this paper would arguably comprise a major achievement in the Majorana problem and topological physics more generally.  Fusion-rule detection promises to reveal the fundamental, in fact \emph{defining}, property of non-Abelian anyons that underlies their exotic exchange statistics.  The first validation of a prototype topological qubit would partially verify the basic tenets of topological quantum computation and thus mark a significant step towards quantum information applications.  An unambiguous demonstration of non-Abelian braiding would march the field further in this direction---while also establishing the most exotic form of exchange statistics permitted by nature.  For all of these milestones we have endeavored to devise controlled experimental protocols that reference non-topological setups mimic only in quite pathological cases.

We focused on one platform---wires coupled to mesoscopic superconducting islands with gate-tunable Coulomb interaction effects---that appears particularly amenable to experimental advances along these lines.  The ability to electrically tune between Coulomb-dominated and Josephson-dominated regimes for the islands indeed enables efficient initialization, manipulation, and readout of Majorana-based qubits through a variety of techniques. Preliminary experiments that seek to optimize gate control over Coulomb effects would be extremely interesting in their own right.  This approach essentially borrows tools from the well-developed area of spin qubits to accelerate progress in the burgeoning field of topological superconductivity.  The philosophy closely resembles the merger of transmons with Majorana platforms advocated in Refs.~\onlinecite{TopTransmon,BraidingWithoutTransport,BeenakkerBraiding}; see also \cite{Ginossar,Yavilberg}.

Our emphasis notwithstanding, the above milestones should apply broadly to Majorana-supporting media and need not rely on the precise platform and manipulation/readout schemes exploited here.  For instance, with only straightforward modifications fusion-rule detection and topological qubit validation experiments should be adaptable to transmon-based devices; other setups such as ferromagnetic atomic chains on superconductors \cite{Yazdani,Nadj-Perge} may require more dramatic modifications and thus pose interesting challenges for future research.  These pre-braiding Majorana milestones are worth pursuing generally since they yield nontrivial physics that belies the simplicity of setups required.

It is of course also interesting to ponder longer-term directions related to our study.  A well-known issue in Majorana systems is that braiding alone does not provide a universal gate set. One way to achieve computational universality involves introducing an additional (unprotected) single-qubit phase gate together with readout of the state encoded by quartets of Majorana modes; for an excellent recent discussion see \cite{NayakReview}.  One could calibrate the former using precisely the same methods that we outlined for topological qubit validation.  Quantifying the fidelity of such a phase gate would constitute a worthwhile experiment complementary to that proposed in Sec.~\ref{QubitSec}.  We expect that readout of Majorana quartets can proceed by performing charge measurements of islands connected center-to-center with a tunable Josephson junction.  And finally, exploring fusion-rule detection and validation of topological qubits remains a fascinating challenge for more exotic types of non-Abelian anyons as well.

\section*{Acknowledgments}
We thank Sven Albrecht, Parsa Bonderson, Michael Freedman, Fabian Hassler, Takis Kontos, Ferdinand Kuemmeth, Olivier Landon-Cardinal, Roman Lutchyn, Karen Michaeli, Roger Mong, Felix von Oppen, Yuval Oreg, Nick Read, and Zhenghan Wang for illuminating discussions.  We acknowledge support from Microsoft Research, the National Science Foundation through grant DMR-1341822 (J.\ A.); the Alfred P.\ Sloan Foundation (J.\ A.); the Caltech Institute for Quantum Information and Matter, an NSF Physics Frontiers Center with support of the Gordon and Betty Moore Foundation through Grant GBMF1250; the Walter Burke Institute for Theoretical Physics at Caltech; the NSERC PGSD program (D.~A.); the Crafoord Foundation (M.~L.\ and M.~H.) and the Swedish Research Council (M.~L.); The Danish National Research Foundation, and the Villum Foundation (C.~M.), The Danish Council for Independent Research $|$ Natural Sciences, and Danmarks Nationalbank (J.~F.). Part of this work was performed at the Aspen Center for Physics, which is supported by National Science Foundation grant PHY-1066293 (R.\ V.\ M.).

\appendix

\newcommand{\FmoveL}{\mathord{\vcenter{\hbox{\includegraphics[scale=1.6]{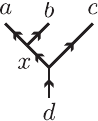}}}}}
\newcommand{\FmoveR}{\mathord{\vcenter{\hbox{\includegraphics[scale=1.6]{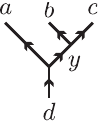}}}}}
\newcommand{\Init}{\mathord{\vcenter{\hbox{\includegraphics[scale=1.6]{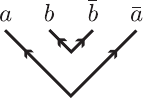}}}}}
\newcommand{\Readout}{\mathord{\vcenter{\hbox{\includegraphics[scale=1.6]{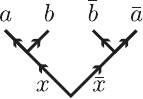}}}}}
\newcommand{\Readouta}{\mathord{\vcenter{\hbox{\includegraphics[scale=1.6]{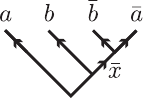}}}}}
\newcommand{\Idab}{\mathord{\vcenter{\hbox{\includegraphics[scale=1.6]{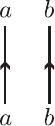}}}}}
\newcommand{\Idabx}{\mathord{\vcenter{\hbox{\includegraphics[scale=1.6]{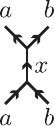}}}}}

\section{Anyon perspective of fusion-rule experiment}
\label{AnyonAppendix}

This Appendix is dedicated to interpreting the fusion-rule protocols from Sec.~\ref{FusionRuleSec} in terms of a general anyon model.
Consider an anyon theory for which particles live at points in two spatial dimensions.
Each particle carries a label indicating the anyon type; for example, in the Ising theory these labels are $I$, $\sigma$, and $\psi$.  
Consistency requires that bringing two point particles together results in another point particle---a process known as \emph{fusion}.
Possible fusion outcomes of particles with labels $a$ and $b$ are described by the fusion rule
\begin{align}
a \times b = \sum_c N^c_{ab} c.
\label{Fusion-Rules}
\end{align}
Here $N_{ab}^c$ are integers denoting the number of ways in which $a$ and $b$ can fuse to $c$.  For simplicity we will assume that all the $N_{ab}^c$ are either $0$ or $1$ (as is the case with the Ising theory).  

All physical processes in the anyon model can be represented diagrammatically.
The diagrams are given by world lines that track the particles' positions as a function of time, with each world line labeled by the corresponding anyon type. 
States are represented by terminating the particle world lines at some constant time slice.
A fusion process corresponds to a trivalent vertex in a diagram, at which two particles coalesce into a third;
the labeled vertices must be compatible with the fusion rules given in Eq.~\eqref{Fusion-Rules}.
And a closed diagram represents an amplitude for a given process.

Many processes can generate the same particles at identical positions---yielding states that live in the same vector space and are hence related by a unitary transformation.  We encountered one example in Fig.~\ref{FusionRuleFig}, where the right path initialized four Ising anyons into the state $|0_{12},0_{34}\rangle$ while the left path initialized the same anyons into the state $\ket{0_{14},0_{23}} = (\ket{0_{12},0_{34}} + \ket{1_{12},1_{34}})/\sqrt2$ [recall Eq.~\eqref{EntangledState}].  The latter expression reflects the non-trivial linear relations between quantum states in different `fusion channels'.
(For an explicit computation of the linear relations see for example Sec.~III~B~2 of Ref.~\onlinecite{TQCreview}).

Diagrammatically, these unitary transformations are given by linear relations on the diagrams. 
The two fundamental relations are so-called $R$- and $F$-moves.  $R$-moves relate diagrams with braiding to those without, while $F$-moves linearly relate the state found from a single particle splitting into three particles in two distinct ways:
\begin{align}
\FmoveL = \sum_{y} \Big( F^{abc}_d \Big)_{xy} \FmoveR.
\label{FmoveDiagramm}
\end{align}
The diagram on the left shows one $d$ particle splitting into an $x$ and $c$ particle, followed by $x$ splitting into $a$ and $b$; the one on the right shows a $d$ particle splitting into $y$ and $a$ and then $y$ splitting into $b$ and $c$.  
The matrix $ F^{abc}_d$---known as an $F$-symbol---is unitary for an anyon model.
Self-consistency of the theory strongly constrains the $F$- and $R$-moves.
The mathematical structure governing this theory is known to be a unitary braided modular tensor category; see Refs.~\onlinecite{Honeycomb, ParsaThesis} for excellent expositions of this structure and how it relates to anyons. 

We now discuss the non-trivial fusion-rule protocol (left path of Fig.~\ref{FusionRuleFig}) in the context of a general anyon model.
The diagram corresponding to the initialization appears on the left side of Eq.~\eqref{FusionExpEquation}: 
First we create two particles $a$ and $\bar{a}$ out of the vacuum and fix their position far from one another.
In Fig.~\ref{FusionRuleFig} this step nucleates Majorana zero modes $\gamma_{1,4}$ at the outer ends of the nanowire.
Next we create two more particles $b$ and $\bar{b}$ out of the vacuum in some other region far from the $a$ and $\bar{a}$ particles.
In Fig.~\ref{FusionRuleFig} this corresponds to nucleating the zero modes $\gamma_{2,3}$ located near the nanowire's center.    
The non-trivial fusion rules reveal themselves when we fuse the left and right pair of anyons and measure the charge. 
In terms of diagrams this means we measure in the basis shown on the right side of
\begin{align}
\Init = \sum_x C_x \Readout.
\label{FusionExpEquation}
\end{align}
The norm square of the coefficients $C_x$ give the probability of a particular fusion process.
For our fusion rule protocol we have $a = b = \bar{a} = \bar{b} = \sigma$, and we measure in the basis $x= I,\psi$.
The coefficients $C_x$ are related to the $F$-symbols since
\begin{align}
\Init =  \sum_ {\bar{x}}  \Big( F^{b\bar{b} \bar{a}}_{\bar{a}} \Big)_{I \bar{x}}  \Readouta.
\end{align}
The diagram on the right further relates to Eq.~\eqref{FusionExpEquation} by pulling the $b$ line past the bottom vertex and moving it over so that it fuses with $a$.
Generically, moving a line in this fashion results only in an extra phase \footnote{The line bending is given by the so-called $A$- and $B$-moves which can be found from the $F$-move.}.  We therefore obtain $|C_x|^2 = \left| \Big( F^{b\bar{b} \bar{a}}_{\bar{a}} \Big)_{I \bar{x}}  \right|^2$. When $a = b = \bar{a} = \bar{b} = \sigma$ then $x = I$ or $\psi$ and $|C_x|^2 = 1/2$.  In other words, the equal measurement probabilities for even- and odd-charge states in the nontrivial fusion-rule protocol, deduced from a different viewpoint in Sec.~\ref{FusionRuleSec}, can be interpreted as a direct measurement of the absolute value of certain $F$-symbols in the Ising theory.

Since the measurement reads off only the absolute value of these $F$-symbols, we can re-interpret this experiment in a different, somewhat more fundamental way.  
The particular matrix elements it measures can alternatively be inferred from the relation
\begin{align}
\Idab = \sum_x \sqrt{\frac{d_x}{d_a d_b}} \Idabx.
\label{Recoupling}
\end{align}
Here $d_x$ is the so-called quantum dimensions of particle type $x$---which can be inferred directly from the fusion rules of the theory. The quantum dimension $d_x$ is the dominant eigenvalue of the matrix $[M(x)]_{ab} = N^a_{xb}$, and physically represents the Hilbert space dimension per particle.  Using Eq.~\eqref{Recoupling} one can predict the probabilities associated with charge measurements on each island, which are given by
\begin{align}
|C_x|^2 &= |C_{\bar{x}}|^2 = \frac{d_x}{d_a d_b} N_{ab}^x,
\end{align}
knowing only the fusion rules and no other information from the anyon theory.  
Inserting the known quantum dimensions for the Ising theory, $d_{I} = d_\psi  = 1$ and $d_{\sigma} = \sqrt{2}$, we recover equal measurement probabilities of 1/2 for each fusion channel.  
In this sense our proposed experiment indeed directly probes the Ising-anyon fusion rules.

\section{Level structure and charge screening by superconducting junctions: Role of barrier transmission}
\label{sec:junction-compare}

Within our approach manipulation of topological superconductors and Majorana zero modes relies critically on efficient gate control over Josephson junctions bridged by nanowires (see, e.g., Fig.~\ref{TwoIslandSetupFig}).  Josephson coupling in these setups has a slightly different physical origin compared to conventional SIS junctions typical for superconducting qubits \cite{Koch}.  Junctions in the latter context host many channels with low transmission probability, so that Josephson coupling can be well modeled with a simple
\begin{equation}
  V_{\rm SIS}(\hat\varphi) = -E_J \cos\hat\varphi
\end{equation}
potential.  We expect that gating in our nanowire devices instead opens a few channels tunable to the regime of large transmission probability (cf.~Refs.~\onlinecite{Gatemon,LangeNanowireJosephson}).  Consider the single junction shown in Fig.~\ref{SetupFig}, which we modeled in Sec.~\ref{GateSec} with a phenomenological Hamiltonian of the form
\begin{eqnarray}
   H & = & E_C \left( \hat{n} - n_0 \right)^2 + V(\hat\varphi), \label{ECplusV} \\
   V(\hat\varphi)& =& -N \Delta \sqrt{1 - T \sin^2 \left( \hat{\varphi} / 2 \right)}, \label{SineFormula}
\end{eqnarray}
where in the second line we have for simplicity assumed all $N$ channels have identical transmission probabilities $T_i = T$.  The goal of this appendix is to contrast the low-lying level structure and screening effects obtained with the standard $V_{\rm SIS}$ potential and that in Eq.~\eqref{SineFormula}.  We focus on the Josephson-dominated regime throughout.

\begin{figure}
\includegraphics[width=\columnwidth]{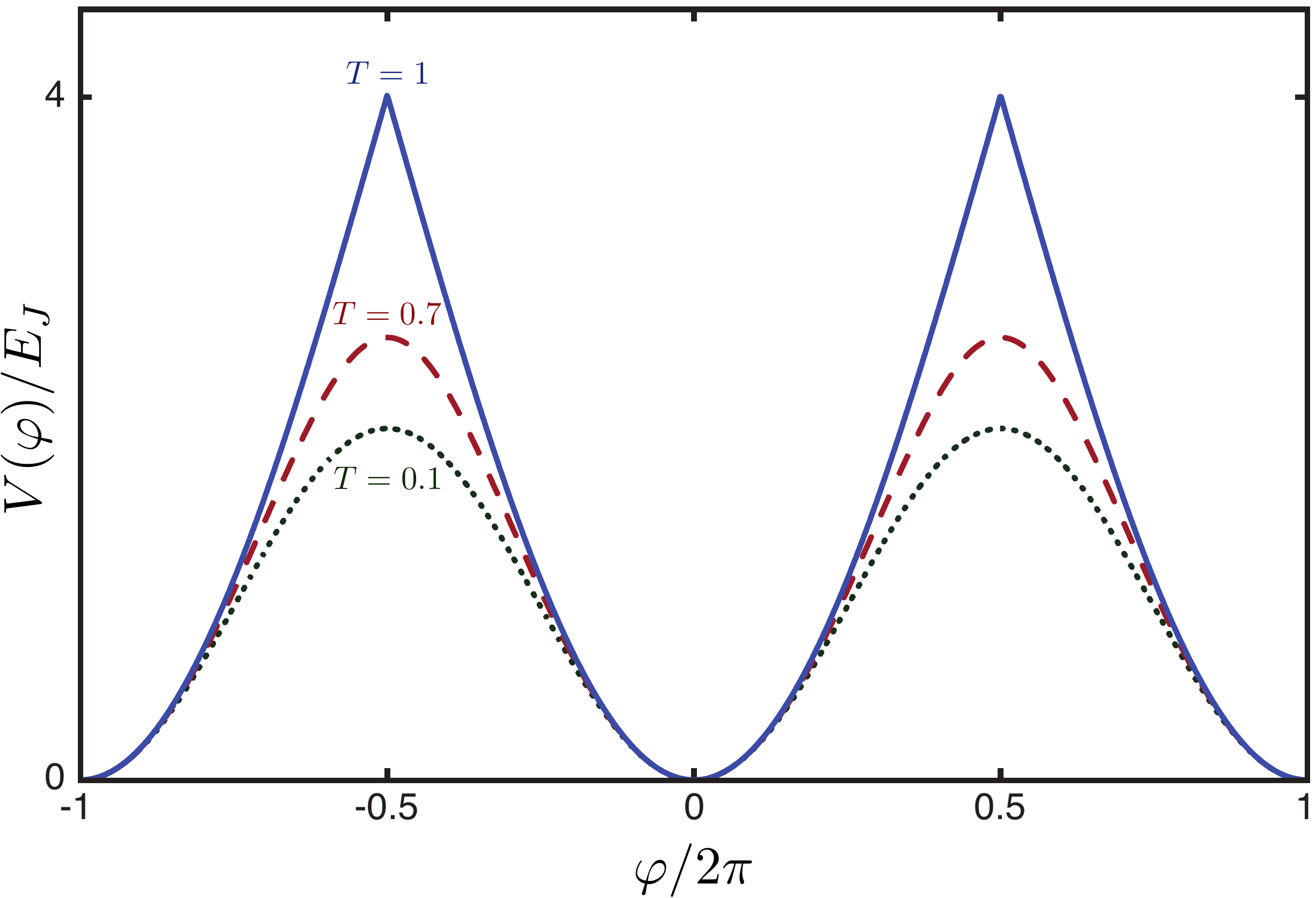}
\caption{Potential $V(\varphi)$ from Eq.~\eqref{SineFormula}, normalized by $E_J$ defined in Eq.~\eqref{EJ_T}, at three different transmissions $T$.  At $T \ll 1$ the potential recovers the usual cosine form familiar from standard SIS junctions, while larger $T$ stiffens the potential compared to this case.  Such stiffening yields relatively minor qualitative effects on the low-lying level structure and charge screening for the junction, even at perfect transmission where $T = 1$. }
\label{PotentialsFig}
\end{figure}

As noted in the main text, Eq.~\eqref{SineFormula} in fact reduces to $V_{\rm SIS}$ in the $T \ll 1$ limit (up to a constant) with
\begin{equation}
  E_J = N\Delta T/4.
  \label{EJ_T}
\end{equation}
For large transmissivity, however, quantitative deviations in the energy spectrum clearly arise for the two cases.  One can anticipate some nontrivial $T$-dependence from the plots of $V(\varphi)$ (normalized by $E_J$) shown in Fig.~\ref{PotentialsFig}, which illustrate `stiffening' of the potential at larger $T$.  (Equivalently, larger $T$ potentials are softened, relative to a parabolic extrapolation, to a lesser degree than a simple cosine.)  We argue that these differences are nevertheless qualitatively rather minor for our purposes and can be reasonably neglected given the rough phenomenological modeling that we employ.

First, we stress that we are interested exclusively in ground states and low-lying excitations.  These low-energy states can be well captured by taking a harmonic approximation for Eq.~\eqref{SineFormula} and expanding about $\hat \varphi = 0$:
\begin{equation}
V(\hat\varphi) \approx \frac{E_J}{2}\hat\varphi^2.
\end{equation}
This is precisely the same small-$\hat\varphi$ dependence obtained by expanding $V_{\rm SIS}$.  Within this rough approximation the two potentials thus yield identical harmonic-oscillator-like spectra with energy levels
\begin{equation}
  E_n = \sqrt{8 E_J E_C}(n+1/2),~~~~~ n = 0,1,2,\ldots
\end{equation}
as discussed in Ref.~\onlinecite{Koch} and sketched in Fig.~\ref{EnergyLevels}(a).  The coupling $E_J$ of course varies with transmission $T$ through Eq.~\eqref{EJ_T}, but splitting between the ground state and first excited state is simply given to a good approximation by $\sqrt{8E_J E_C}$ for any $T$ between 0 and 1.  We have verified this behavior by simulating the Hamiltonian with the exact potentials numerically.

In the harmonic treatment above, the ground state exhibits an \emph{exact} two-fold degeneracy consisting of even- and odd-parity states; moreover, the energies are independent of the gate-tuned offset charge $n_0$ in Eq.~\eqref{ECplusV}.   We turn now to the full anharmonic potentials.  Anharmonicity both lifts the exact degeneracy and causes the energies to disperse with $n_0$ as depicted in Fig.~\ref{EnergyLevels}(a).
Let us focus on the even-parity ground-state sector and consider the energy difference
\begin{equation}
  \Delta E_{\rm max} \equiv \max E_0(n_0) - \min E_0(n_0)
\end{equation}
between the maximal and minimal ground-state energies $E_0(n_0)$ obtained upon varying the offset $n_0$.  Note that $\Delta E_{\rm max}$ is equivalent to the maximal degeneracy splitting between even- and odd-parity states [recall Fig.~\ref{EnergyLevels}(a)] and is thus particularly important for this work.

The energy $\Delta E_{\rm max}$ is closely connected with the probability for phase slips $\varphi \rightarrow \varphi + 2 \pi$ that tunnel through maxima of the potential barrier in Fig.~\ref{PotentialsFig} \cite{NazarovBook}.  Since increasing $T$ `stiffens' the confining potential (in units of $E_J$) we expect the maximal degeneracy splitting to be parametrically further suppressed as $T\rightarrow 1$.  For the Josephson-dominated limit $E_J/E_C \gg 1$ considered here, we can compute $\Delta E_{\rm max}$ for general $T$ using the same WKB approach \cite{Connor84Eigenvalues} as for the cosine potential $V_{\rm SIS}$ valid at $T \ll 1$ \cite{Koch}.  The result takes the form
\begin{equation}
  \Delta E_{\rm max} = \alpha(T) E_C\left( \frac{E_J}{E_C} \right)^{3 / 4} e^{- \beta(T) \sqrt{8 E_J / E_C}}.
  \label{eq:epst1}
\end{equation}
For $T \ll 1$ we have $\beta=1$ \cite{Koch}, whereas in the opposite limit of $T=1$ we find a mild enhancement to $\beta \approx 1.17$; the prefactor $\alpha$ is also slightly reduced for increased $T$.  Direct numerical simulations of Eqs.~\eqref{ECplusV} and \eqref{SineFormula} validate this analysis (for simulation details see Ref.~\onlinecite{Hell}). Increasing $T$ at fixed $E_J$ thus reduces the degeneracy splitting, consistent with the above intuition.  This observation justifies our use of the simpler potential $V_{\rm SIS}$ to conservatively estimate the degeneracy splittings and related time scales in Sec.~\ref{RealisticCase}.

Despite the quantitative differences highlighted above, suppression of charging effects is qualitatively the same for a junction with either many channels of low transmission or a few channels with large transmission. Overall, the physics is that, at least within the model used here, the superconductor cannot screen charge perfectly because of its gap $(E_J \sim \Delta)$, in stark contrast to a normal conducting Fermi sea.  It should however be noted that this analysis does not include all renormalization effects close to perfect transmission---see also Ref.~\onlinecite{Feigelman_ECrenorm}.

It would be quite interesting in future work to analyze the degeneracy splittings, and junction properties more generally, in a microscopic description that more faithfully captures details of our mesoscopic-island setups. It remains an open question how to describe the junction in the limit where the charging energy $E_C$ is not small compared to $\Delta$; the Josephson energy may then cross $E_C$ when the transmission probabilities are not small, in which case the charging energy can no longer be treated as a perturbation.

\section{Topological qubit relaxation time $T_1$}
\label{T1appendix}

To estimate the topological qubit's $T_1$ time we will invoke several simplifications.  First, we continue to use the two-level qubit Hamiltonian in Eq.~\eqref{Hqubit} to treat fluctuations that produce relaxation.  Thus we incorporate only \emph{classical} noise.  This assumption confines our analysis of $T_1$ to the regime where the qubit frequency $\omega_0$ is small compared to temperature \cite{Schoelkopf02_arXiv_0210247}, which is actually quite reasonable given its exponential suppression.  (One should at least expect $\hbar\omega_0 \lesssim k_B T$ experimentally; see Sec.~\ref{RealisticCase} and Ref.~\cite{Marcus15_NatureNano_10_232}.)  Second, we neglect fluctuations in the $\sigma^z$ coupling and simply set $h_z(t) = \hbar \omega_0/2$ since fluctuations in $h_z$ are more important for dephasing than relaxation.  Third, we retain only the random part of the $\sigma^x$ coupling, taking $h_x(t) = \delta h_x(t)$ where $\delta h_x(t)$ satisfies Gaussian correlations
\begin{equation}
  \langle \delta h_x(t) \rangle = 0,~~~~\langle \delta h_x(t) \delta h_x(t')\rangle = S_x(t-t').
\end{equation}
Assuming again $1/f$ noise, we take
\begin{equation}
  S_x(\omega) \sim \frac{(\Delta E_{\rm typ}^x)^2}{|\omega|}
  \label{xNoisePower}
\end{equation}
with $\Delta E_{\rm typ}^x$ the typical fluctuation amplitude in the $\sigma^x$ coupling.  The uniform part of $h_x(t)$ is neglected since it only produces oscillations on top of the decaying signal obtained in Eq.~\eqref{PforT1} below.  And finally, we treat the noise perturbatively, i.e., $\Delta E_{\rm typ}^x \ll \hbar \omega_0$---a quite reasonable requirement when the middle valve in Fig.~\ref{QubitFig}(a) is maximally closed.

With these simplifications, time-dependent perturbation theory \cite{Schoelkopf02_arXiv_0210247} gives the following noise-averaged transition rates (probability per unit time) between the $|0\rangle$ and $|1\rangle$ qubit states:
\begin{equation}
  \Gamma_{1\rightarrow0} = \Gamma_{0\rightarrow1} =\frac{S_x(\omega_0)}{\hbar^2}.
\end{equation}
Note that equality of the two transition rates follows because the noise is classical and thus satisfies $S_x(\omega) = S_x(-\omega)$ [which reality of $\delta h_x(t)$ guarantees].  The equilibrium populations $w_{0,1}$ of $|0\rangle$ and $|1\rangle$ are then equal.
Indeed, invoking detailed balance relates these populations to the transition rates according to
\begin{equation}
\frac{w_0}{w_1} = \frac{\Gamma_{1\rightarrow0}}{\Gamma_{0\rightarrow1}} = 1.
\label{ratios}
\end{equation}
This is why our classical-noise assumption effectively restricts our treatment of relaxation to the limit $\hbar \omega_0 \ll k_B T$ as remarked earlier.

Suppose now that one prepares the qubit into state $|0\rangle$ at time $t = 0$. The probability of measuring the qubit in state $|1\rangle$ at a later time $t$ is then given by
\begin{equation}
\langle \tilde P(t)\rangle = \frac{1}{2}\left(1 - e^{-t/T_1} \right)
\label{PforT1}
\end{equation}
with relaxation time~\cite{Schoelkopf02_arXiv_0210247}
\begin{equation}
T_1 = \frac{1}{2}\frac{1}{\Gamma_{0\rightarrow1}} = \frac{1}{2}\frac{\hbar^2}{S_x(\omega_0)} \sim \frac{\hbar^2 \omega_0}{(\Delta E_{\rm typ}^x)^2}.
\label{T1eq}
\end{equation}
Thus we have obtained the $\omega_0$-dependence quoted in the main text.  The connection between fluctuations and time-averaged quantities for the topological qubit correspondingly establishes the unconventional scaling relation between $T_1$ and the dephasing time $T_2$ provided in Eq.~\eqref{scaling2}.

Note that as with $T_2$, the precise form of $T_1$ clearly depends on details of the noise model.  Nevertheless, some nontrivial dependence on $\omega_0$ is all that we require to nontrivially link the qubit coherence times.  The dependence of $T_1$ on $\Delta E_{\rm typ}^x$, on the other hand, is more straightforward:  $1/T_1 \propto (\Delta E_{\rm typ}^x)^2$ independent of noise model, by the very general Eq.~\eqref{T1eq}.  According to Eq.~\eqref{Eii} we expect $\Delta E_{\rm typ}^x \propto e^{-W/\xi_U}$, where $\xi_U$ is the decay length of the Majorana wavefunctions into the region of width $W$ separating the islands. One can tune $\xi_U$ by modulating the middle valve in Fig.~\ref{QubitFig}(a) to experimentally probe the dependence of $T_1$ on $\Delta E_{\rm typ}^x$, which would certainly be worthwhile as a further quantitative qubit characterization.  We caution, however, that such an experiment does not by itself provide evidence for topological protection.  Indeed, the same exponential dependence predicted for $\Delta E_{\rm typ}^x$ would appear for the conventional Andreev qubit in Fig.~\ref{QubitFig}(b).  Sharp distinctions instead follow through the dependence on $\omega_0$ noted above and in the main text.

Generally speaking, our treatment of qubit decoherence times $T_1$ and $T_2$ is meant to highlight universal aspects of the physics, as is the theme for much of the present work.  We leave for future work a more microscopic treatment of relaxation processes relevant to our putative topological qubit, including consideration of quantum mechanical properties of specific noise sources.

\bibliography{Majorana_milestones}

\end{document}